\documentclass[useams,usenatbib]{mn2e}
\usepackage{times}
\usepackage{graphicx}

\title[NGC~1569-B]{Anatomy of a Young Massive Star Cluster: NGC~1569-B
\thanks{Based in part on data obtained at the W.M.Keck Observatory,
        which is operated as a scientific partnership among the California
        Institute of Technology, the University of California, and the National
        Aeronautics and Space Administration. The Observatory was made possible
        by the generous financial support of the W.M. Keck Foundation.
	Also based on observations with the NASA/ESA Hubble Space Telescope,
	obtained from the data archive at the Space Telescope Science
	Institute. STScI is operated by the association of Universities
	for Research in Astronomy, Inc.\ under the NASA contract
	NAS 5-26555.}
}

\author[Larsen et al.]{S. S. Larsen$^1$, L. Origlia$^2$, J. Brodie$^3$,
  J. S. Gallagher, III$^4$.\\
 $^1$ Astronomical Institute, University of Utrecht, Princetonplein 5,
     NL-3584 CC Utrecht, The Netherlands, e-mail larsen@astro.uu.nl \\
 $^2$ INAF-Osservatorio Astronomico di Bologna, Via Ranzani 1, I-40127 Bologna,
      Italy, e-mail livia.origlia@bo.astro.it \\
 $^3$ UCO/Lick Observatory, 1156 High Street, University of California, 
      Santa Cruz, CA 95064, USA, email brodie@ucolick.org \\
 $^4$ Astronomy Department, University of Wisconsin, 475 North Charter
      Street, Madison, WI 53706, USA, email jsg@astro.wisc.edu
       }
\date{\today}

\begin{document}
\pagerange{\pageref{firstpage}--\pageref{lastpage}} \pubyear{2007}
\maketitle
\label{firstpage}

\begin{abstract}
We present new $H$-band echelle spectra, obtained with the NIRSPEC
spectrograph at Keck II, for the massive star cluster ``B''
in the nearby dwarf irregular galaxy NGC~1569. From spectral synthesis and 
equivalent width measurements we obtain abundances and abundance patterns.
We derive an Fe abundance of [Fe/H]=$-0.63\pm0.08$,
a super-solar [$\alpha$/Fe] abundance ratio of $+0.31\pm0.09$, and
an O abundance of [O/H]=$-0.29\pm0.07$.
We also measure a low  $\rm ^{12}C/^{13}C\approx 5\pm1$ isotopic ratio.
Using archival imaging from the Advanced Camera for
Surveys on board HST, we construct a colour-magnitude diagram (CMD) for the
cluster in which we identify about 60 red supergiant (RSG) stars, consistent 
with the strong RSG features seen in the $H$-band spectrum. The mean
effective temperature of these RSGs, derived from their observed colours and
weighted by their estimated $H$-band luminosities, is 
3790 K, in excellent agreement with our spectroscopic estimate of
$T_{\rm eff} = 3800\pm200$ K. From the CMD we derive an age of 15--25 Myr, 
slightly older than previous estimates based on integrated broad-band colours.
We derive a radial velocity of $\langle v_r \rangle=-78\pm 3$ km/s
and a velocity dispersion of $9.6\pm0.3$ km/s. In combination with an
estimate of the half-light radius of $0\farcs20\pm0\farcs05$ from the HST 
data, this leads to a dynamical
mass of $(4.4\pm1.1)\times10^5$ M$_\odot$.  The dynamical mass
agrees very well with the mass predicted by simple stellar population
models for a cluster of this age and luminosity, assuming a normal
stellar IMF. The cluster core radius appears smaller at longer wavelengths,
as has previously been found in other extragalactic young star clusters.

\end{abstract}

 
\section{Introduction}

Massive star clusters are potentially useful test particles for 
constraining the evolutionary histories of their host galaxies. 
They can remain observable for the entire lifetime of a galaxy
(as illustrated by the old \emph{globular clusters} which surround 
every major galaxy), and they are bright enough to be studied in detail
well beyond the Local Group. The internal velocity broadening of their
spectra is typically only a few km/s, making detailed abundance analysis
at high spectral resolution feasible. In a previous paper we have taken
a first step towards exploiting this potential by analysing $H$ and $K$-band 
spectra of a young massive star cluster (YMC) in the nearby spiral galaxy 
NGC~6946 \citep{lar06}. Here we apply a similar analysis to one of the 
young star clusters in the nearby (post-) starburst galaxy NGC~1569. 

NGC~1569 was one of the first galaxies in which the presence of exceptionally
bright, young star clusters was suspected. A spectrum of one of
the two bright ``stellar condensations'' in NGC~1569 was obtained already by 
\citet{mayall35}, although the true nature of these objects was probably
first discussed in detail by \citet{as85}.  They found that 
the spectra are of composite nature, and also noted that observations 
with the Hubble Space Telescope (HST)
would definitively settle the issue whether or not these objects are 
really ``super star clusters''.  Indeed, pre-refurbishment mission HST 
observations by \citet{oco94} settled the issue by showing that both objects 
are extended,
with half-light radii of about two pc. Based on high-dispersion spectroscopy
from the HIRES spectrograph on the Keck~I telescope, a dynamical mass
of $\approx 10^6$ M$_\odot$ was soon after estimated for the brighter of the
two objects, NGC~1569-A \citep{hf96,stern98}. 

The clusters in NGC~1569 are prime candidates for abundance analysis
in the near-infrared where they are very bright. The IR holds a particular
advantage over optical studies for NGC~1569 due to the large amount of
foreground extinction.
\citet{gg02} found the integrated $H$-band 
spectra of both clusters A and B to be well approximated by red supergiant 
templates.  Although NGC~1569-A is the brighter of the two, it
is actually a binary cluster itself with some evidence for a (small) age 
difference between the two components \citep{guido97,maoz01,origlia2001}. 
Whether or not the two components are physically connected or the result
of a chance projection is unknown.  In this paper we concentrate 
on NGC~1569-B.  

In addition to providing insight into the histories of their host galaxies,
young star clusters with masses in excess of $10^5$ M$_\odot$ also offer an
excellent opportunity to study large samples of coeval massive stars. Such 
clusters are rare in the Milky Way and even in the Local Group, so it is 
necessary to extend the search to a larger volume.  The most massive known 
young star clusters in the Milky Way disk are Westerlund 1 \citep{clark05} 
and an object discovered in the 2MASS survey \citep{figer06}, both of which 
may have masses approaching $10^5$ M$_\odot$. Young clusters of similar
masses are found in the Large Magellanic Cloud \citep{vdb99}, but these
pale in comparison with objects like NGC~1569-A and NGC~1569-B.

We have obtained new NIRSPEC $H$-band spectra of NGC~1569-B, optimised for 
abundance analysis, and we additionally present photometry for 
\emph{individual} stars in the cluster derived from archival observations 
with the high resolution channel (HRC) of the Advanced Camera
for Surveys (ACS) on HST.  The
HST data provide an independent verification of the stellar parameters
(notably $T_{\rm eff}$ and $\log g$) derived from the spectral analysis, and 
also allow
us to compare the observed colour-magnitude diagram (CMD) with standard 
isochrones.  Finally, 
we use new measurements of the structural parameters and velocity dispersion
to derive the cluster mass and mass-to-light ratio and compare with
predictions by simple stellar population (SSP) models.

We begin by briefly describing the data in \S\ref{sec:data}. The analysis
and our main results then follow in \S\ref{sec:results}, where we 
first discuss the near-infrared spectroscopy and the abundance analysis
(\S\ref{sec:abundance}). We then proceed to construct a CMD
from the HST imaging (\S\ref{sec:cmd}) from which we derive
stellar parameters for the red supergiants (RSGs) that 
are compared against the spectroscopic results (\S\ref{sec:speccmp})
and theoretical isochrones (\S\ref{sec:interp}). In \S\ref{sec:virmass}
we measure structural parameters for NGC~1569-B and combine these with
velocity dispersion measurements to derive a dynamical mass estimate.
Finally, some additional discussion and a summary are 
given in \S\ref{sec:summary}.

\section{Data}
\label{sec:data}

\subsection{NIRSPEC data}

We observed NGC~1569-B in the $H$-band with the NIRSPEC spectrograph
\citep{ml98} on the Keck~II telescope on 2006 November 8, using the high 
resolution echelle mode with a slit width of $0\farcs432$ (3 pixels). This 
yields a spectral resolution of $\lambda/\Delta \lambda = 25 000$. Since
the NIRSPEC detector does not cover the entire $H$-band echellogram in a
single exposure, we optimised the tilts of the NIRSPEC gratings so that
the number of features suitable for abundance determination was maximised.
To facilitate easy sky subtraction, we followed the standard strategy of
obtaining pairs of exposures nodded
a few arcseconds along the slit. A total of 20 integrations, each with an 
exposure time of 300 s, were made, resulting in a total exposure time
of 6000 s. One-dimensional spectra were extracted from each pair of
nodded exposures and wavelength calibrated using the REDSPEC IDL package 
written by L.\ Prato, S.\ S.\ Kim \& I.\ S.\ McLean. The
individual 1-D spectra were then co-added.  The final summed 1-D spectra 
have a S/N of about 100 per pixel, or $\sim175$ per 3 pixel resolution 
element. 

\subsection{HST/ACS data}

We use observations of NGC~1569 obtained with the 
ACS HRC as
part of programme GTO-9300 (P.I.: H.\ Ford). The HRC has a field of
view of about $29\arcsec\times26\arcsec$ and a pixel scale of 
$\sim0\farcs027$ pixel$^{-1}$, but with significant variations across 
the field due to the geometric distortions in ACS. For the photometry
of individual stars we use the F555W ($\approx$ V-band) and F814W 
($\approx$ I-band) images, each exposed 
for $3\times130$ s. For measurements of structural parameters we
also include F330W ($\approx$ U-band) data ($3\times220$ s).  The data were
downloaded from the HST archive at the \emph{Space Telescope - European
Coordinating Facility} (ST-ECF) and individual exposures were
co-added with the MULTIDRIZZLE task \citep{koek02}.
Multidrizzle filters out cosmic ray hits and detector defects,
corrects for geometric distortions and projects
the images on an orthogonal pixel grid with a scale of 
exactly $0\farcs025$  pixel$^{-1}$.
At a distance of 2.2 Mpc (\S \ref{sec:dist}), this corresponds to 
a linear scale of 0.27 pc pixel$^{-1}$.

\subsection{Distance and Reddening of NGC~1569}
\label{sec:dist}

While it is clear from the high degree of resolution into individual
stars that NGC~1569 must be nearby, the exact distance remains quite uncertain. 
One uncertainty is in the correction for foreground extinction, for which 
the NASA Extragalactic Database (NED) gives either $A_B = 2.03$ mag 
\citep{bh82} or $A_B = 3.02$ mag \citep{sch98}. \citet{is88} estimated 
$E(B-V) = 0.56\pm0.10$ or $A_B = 2.30\pm0.40$ mag (for $R_B = 4.1$).  
Combined with the apparent distance modulus $(m-M)_B = 29.0$ derived by 
\citet{as85}, this leads to $(m-M)_0 = 26.7\pm0.6$ or $D=2.2\pm0.6$ Mpc, 
which is the value used by \citet{origlia2001} and several other studies
in the literature. Based on photometry for a small number of 
yellow and blue supergiants and also using $E(B-V) = 0.56$, \citet{oco94} 
estimated $D = 2.5\pm0.5$ Mpc. Using observations of the red giant branch 
tip, \citet{mk03} derived two possible distances of either $D=1.95\pm0.2$ Mpc
or $D=2.8\pm0.2$ Mpc, both assuming the Schlegel et al.\ foreground 
extinction.  For the lower \citet{is88} extinction, the distances would
instead be $D=2.7\pm0.3$ Mpc or $D=3.9\pm0.3$ Mpc.  In this paper we
use the same value of 2.2 Mpc as \citet{origlia2001} and the extinction
value from \citet{is88}, i.e.\ $E(B-V) = 0.56$ or $A_B = 2.30$ mag
($A_V = 1.74$ mag). The abundance analysis is not affected by either
of these parameters. For the analysis of the colour-magnitude diagram
and the cluster mass we will give further comments when needed.

\section{Analysis and Results}
\label{sec:results}

\subsection{Chemical abundances}
\label{sec:abundance}

\begin{table*}
\begin{center}
\caption{Adopted stellar atmosphere parameters and abundance estimates for cluster B in NGC~1569.
\label{ab}}
\begin{tabular}{lccccccccccc}
\hline
$T_{\rm eff}$ [K] &
log~g &
$\xi$ [Km~s$^{-1}$]   &
$\rm [Fe/H]$   &
$\rm [O/Fe]$   &
$\rm [Ca/Fe]$   &
$\rm [Si/Fe]$   &
$\rm [Mg/Fe]$   &
$\rm [Ti/Fe]$   &
$\rm [\alpha/Fe]^a$&
$\rm [Al/Fe]$   &
$\rm [C/Fe]$   \\
\hline
3800 & 0.0 & 3 & $-0.63$ & 0.34 & 0.33 & 0.33 & 0.23 & 0.33 & 0.31 & 0.23 & $-0.27$ \\ 
$\pm200$ & $\pm0.5$ & &$\pm$0.08&$\pm$0.10 & $\pm$0.09&$\pm$0.11&$\pm$0.13&$\pm$0.10&$\pm$0.09&$\pm$0.11&$\pm$0.10\\
\hline
\end{tabular}
\end{center}
$^a$ $\rm [\alpha/Fe]$ is the average $\rm [<Ca,Si,Mg,Ti>/Fe]$ abundance ratio.
\end{table*}

\begin{figure*}
\centering 
\includegraphics[width=15cm]{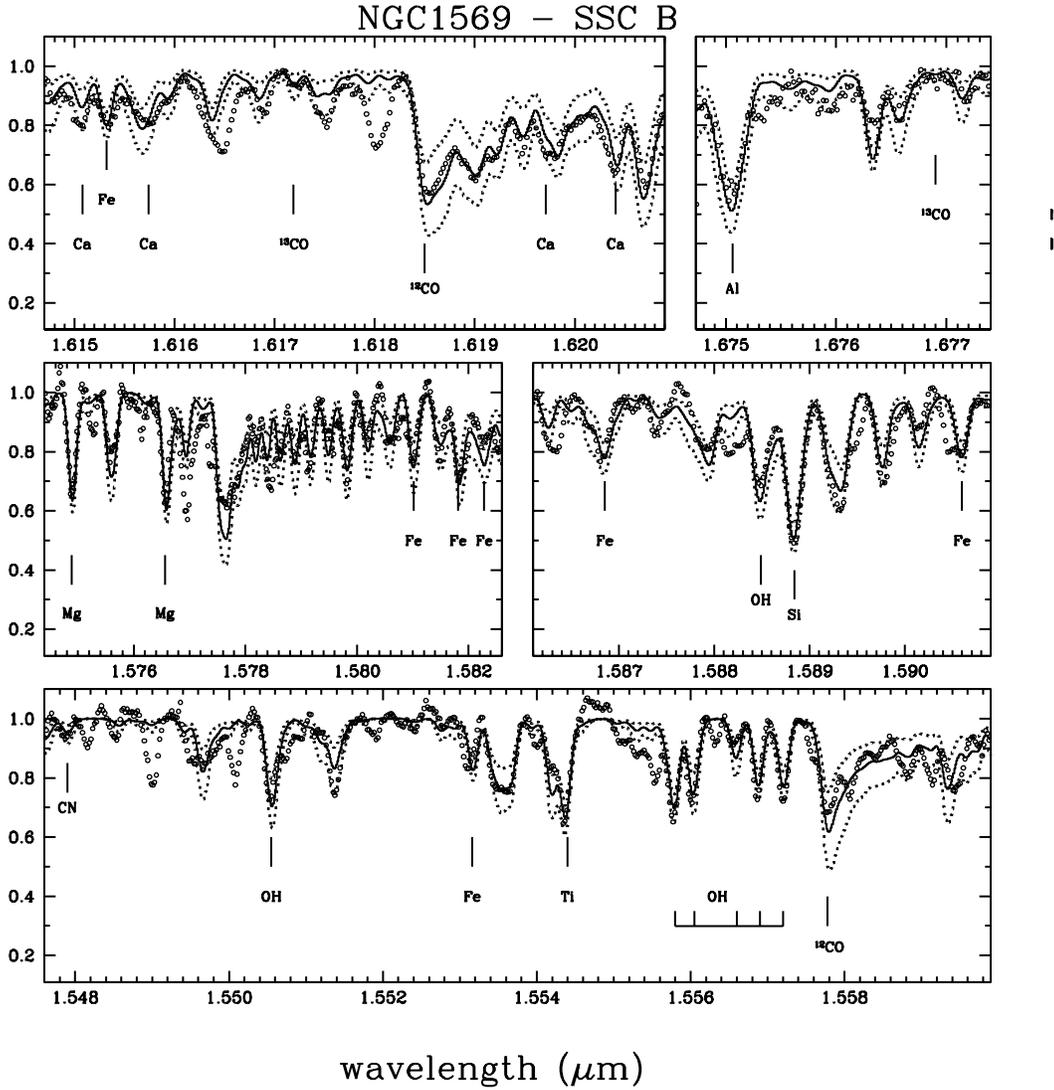}
\caption {\label{fig:spec}Near-IR spectra of the SSC B of NGC~1569.
Observed spectrum: open circles. Solid line: best-fitting RSG spectrum.
Dotted lines: Model spectra with abundances scaled by a factor of 0.5 and 2.0.
A few atomic and molecular features of interest are also marked.
\label{IR}}
\end{figure*}

In general, abundance analysis from integrated spectra requires full spectral 
synthesis techniques to properly account for line blending and population 
synthesis to define the dominant contribution to the stellar luminosity.
The near IR stellar continuum of young stellar clusters and starburst galaxies
is almost entirely due to luminous RSGs
\citep[see e.g.][for a review]{oo00} and usually it dominates
over nebular and dust emission.
This represents a major, conceptual simplification in population and
spectral synthesis techniques, making the interpretation
of integrated spectra much easier.  The spectra can be modelled with
an equivalent, average star, whose stellar parameters
(temperature $T_{\rm eff}$, gravity log{\it g} and microturbulence
velocity $\xi$) mainly depend on the age and metallicity.

Our analysis of the NIRSPEC spectra of NGC~1569-B closely followed that in 
our study of a YMC in NGC~6946 \citep{lar06}. 
A grid of synthetic RSG spectra for different input
atmospheric parameters and abundances were computed, using an updated
\citep{ori02,ori03} version of the code described in \citet{ori93}.
Briefly, the code uses the LTE approximation and is based on molecular
blanketed model atmospheres of \citet{jbk80} at temperatures $\le$4000~K.
Recently, the NextGen model atmospheres \citep{hau99} have been also
implemented within the code and tested, yielding very similar results.
The code also includes several thousands of near IR atomic lines and molecular
roto-vibrational transitions due to CO, OH and CN.
Three main compilations of
atomic oscillator strengths are used, namely
the Kurucz database\footnote{
http://cfa-www.harward.edu/ \\
amdata/ampdata/kurucz23/sekur.html}
and those published by \citet{bg73} and \citet{mb99}.
For further details we refer 
to \citet{ori93,ori02} and \citet{ori03}.

As was done for the YMC in NGC~6946, we estimated the random errors on
the derived quantities using a number of test models. With respect to
the best-fitting spectrum, we varied the temperature by 
$\Delta T_{\rm eff}$ of $\pm$200K, 
the microturbulence by $\Delta \xi$ of $\mp$1.0 km s$^{-1}$,
gravity by $\Delta \log~g$ of $\pm$0.5 dex and
abundances varying accordingly by $\approx\pm0.2$~dex,
in order to still reproduce the depth of the observed features. 
By comparing the mean difference ($\delta$) between the observed and
best-fitting model spectrum with the corresponding $\delta$-value for 
each test model, we found that the test models are at best statistically 
significant at $\ge 2\sigma $ level, only. See \citet{lar06} for details.

Fig.~\ref{fig:spec} shows several regions of the cluster spectrum with the 
most prominent features indicated.  As noted in several previous studies, the 
spectrum shows many absorption features indicating the presence of RSGs.
The open circles are the observed spectrum,
the solid line the best fitting RSG model spectrum and the dotted lines
are models with a factor of two different abundances from the best fit
solution. Synthetic spectra with lower element abundances are
\emph{systematically} shallower than the best-fit solution, while the 
opposite occurs when higher abundances are adopted.  

By fitting the full observed $H$-band spectrum and by measuring the equivalent
widths of selected lines, we obtained the stellar parameters and
abundance patterns listed in Table~\ref{ab}.  Reference Solar abundances are 
from \citet{grev98}.  Among the measured quantities are an
Fe abundance of [Fe/H] = $-0.63\pm0.08$ and 
[O/H] = $-0.29\pm0.07$ (i.e.\ $12 + \log$ (O/H) = 8.54), which combine to
[O/Fe] = $0.34\pm0.10$.
We also measure an average [$\alpha$/Fe]=$0.31\pm0.09$ enhancement for
the other $\alpha$-elements and $\rm ^{12}C/^{13}C\approx5\pm1$. 
The latter ratio is even lower than
the value we found for the YMC in NGC~6946 (${\rm ^{12}C/^{13}C} = 8\pm2$).
Isotopic $\rm ^{12}C/^{13}C$ abundance ratios below 10 are quite common in 
Galactic RSGs \citep{lt74,tll76,gw00} and have also been found
in the cluster NGC~330 in the Small Magellanic Cloud \citep{gw99}.  
The fact that the observed ${\rm ^{12}C/^{13}C}$ values in RSGs
tend to be lower than
standard stellar models predict \citep{schal92} may indicate that 
extra mixing mechanisms are present.

To our knowledge, no previous determination of the Fe abundance in 
NGC~1569 exists. Our value is similar to that of cool supergiants in 
the Small Magellanic Cloud \citep{hill99}. Our oxygen abundance for
NGC~1569-B, on the other hand, 
is more similar to that of young stars in the Large Magellanic 
Cloud, although the number of stars with reliable abundance determinations is
small \citep{rtd02}. Our O abundance for NGC~1569-B is
about a factor of two higher than the commonly cited value of 
12 + log (O/H) = $8.19\pm0.02$ for H{\sc ii} regions in NGC~1569 \citep{ks97}.
Other measurements of the O abundance in NGC~1569 come from
\citet{sdk94} who derived $12+\log$ (O/H) = 8.37,
intermediate between our value and that of \citet{ks97},
while \citet{dev97} got $12+\log$ (O/H) = 8.26. All these previous 
measurements refer to the ionised gas, however, and we are not aware of
any previous measurement of the O abundance in NGC~1569 from stellar
spectroscopy.

From the spectral fits we also measure a 
heliocentric radial velocity of $\langle v_r \rangle=-78\pm 3$ km/s and 
velocity dispersion (corrected for instrumental broadening) of 
$\approx9.9\pm1$ km/s.  This velocity dispersion is higher than the 
$7.5\pm0.2$ km/s found by \citet{gg02}, but agrees well with our measurement 
based on cross-correlation analysis (\S \ref{sec:vd}). The radial
velocity itself is in excellent agreement with the optical radial
velocity of NGC~1569 listed in the 
\textit{Third Reference Catalogue of Bright Galaxies} 
\citep[$-74\pm17$ km/s;][]{rc3}.

\subsection{The cluster colour-magnitude diagram}
\label{sec:cmd}

\begin{figure}
\centering
\includegraphics[width=40mm]{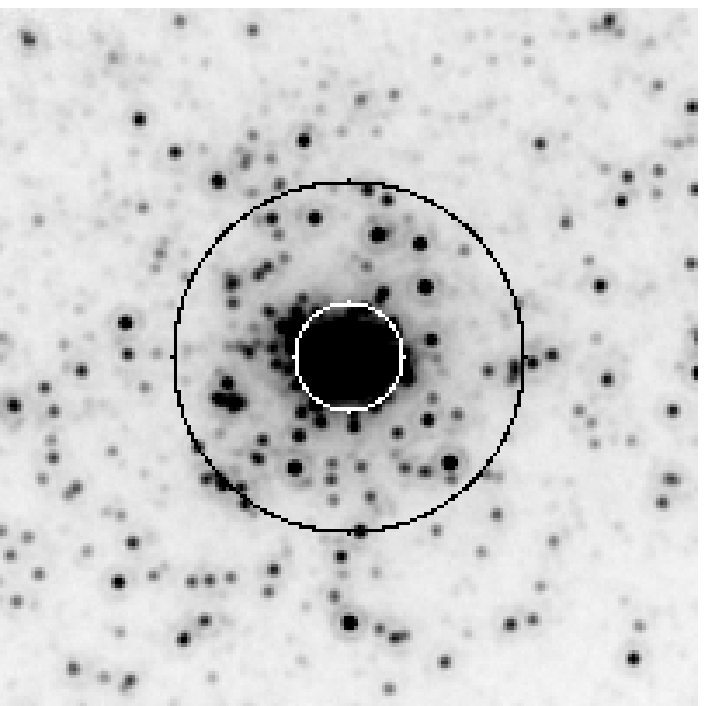}
\includegraphics[width=40mm]{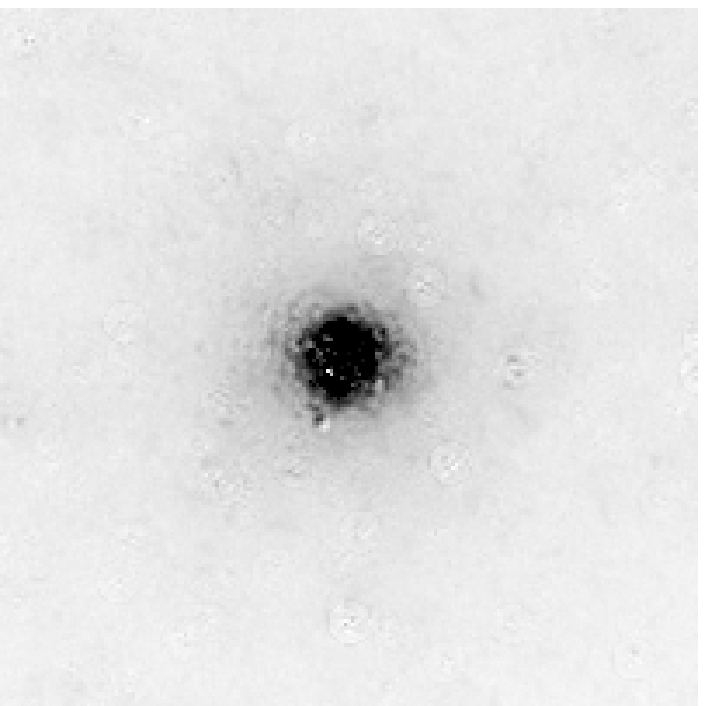}
\caption{\label{fig:phot}Left: ACS/HRC F814W image of NGC~1569-B. The
circles indicate the region for which a CMD was constructed.  Right:
The residual image produced by DAOPHOT/ALLFRAME.
}
\end{figure}

\begin{figure}
\centering
\includegraphics[width=85mm]{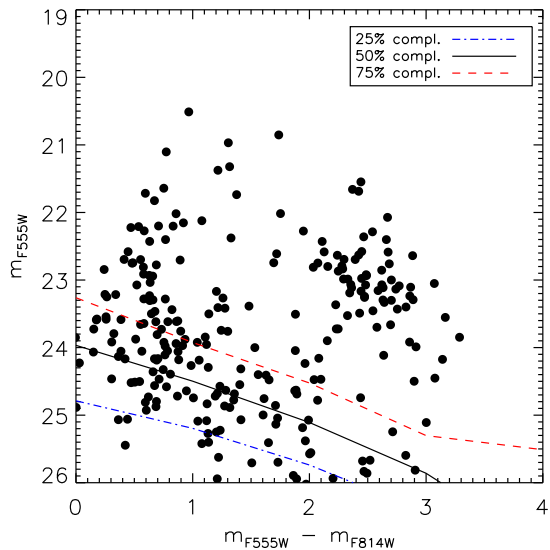}
\caption{\label{fig:cmdraw}
Colour-magnitude diagram for NGC~1569-B. The three curves show
the 75\% (dashed line, red in the on-line edition), 50\% (solid line) and 
25\% (dotted-dashed line, blue in the on-line edition) completeness limits 
based on artificial star experiments.
}
\end{figure}

\subsubsection{Photometry}

The HRC images show NGC~1569-B resolved into its brightest constituent
stars. This presents us with a welcome opportunity to check the properties
of RSGs derived from the spectral fits.  To study the CMD, we therefore 
carried out point-spread-function
(PSF)-fitting photometry with the DAOPHOT III photometry package
\citep{stetson87,stetson90}.
Stars were first detected in the F814W image with the FIND task, aperture 
photometry was obtained for the F555W and F814W images with the PHOT task 
and PSFs were constructed using 7 isolated stars scattered 
across the HRC detector.  A first iteration of PSF-fitting photometry was then 
carried out by running the ALLFRAME task on the F555W and F814W frames
simultaneously.  The FIND routine was applied a 
second time to the star-subtracted F814W image produced by ALLFRAME, and
improved PSFs were constructed
by running the PSF task once again on images where all stars detected in
the first pass (except the PSF stars) had been subtracted.  The ALLFRAME
PSF-fitting 
was then repeated with the combined coordinate list. We found that two such
iterations were sufficient to produce very clean final star-subtracted 
images. 
Fig.~\ref{fig:phot} shows the F814W image
of NGC~1569-B in the left-hand panel and the residual image produced by
the second ALLFRAME iteration in the right-hand panel. The 
stars subtract out
nicely without any systematic residuals, even quite close to the centre
of the cluster, indicating the PSF-fitting procedure went well.  

For calibration to
the VEGAMAG system we used the photometric zero-points for the
HRC published on the ACS world-wide-web-page ($z_{\rm F555W} = 25.255$ mag,
$z_{\rm F814W} = 24.849$ mag).  To fix the zero-points of the
PSF fitting photometry we carried out aperture photometry in a radius
of 5 pixels ($0\farcs125$) for a number of bright stars on images
where all other stars had been subtracted and calculated the offset
between the aperture and PSF photometry for these stars. We estimate that 
the uncertainty
in tying the PSF and aperture photometry together is about 0.01 -- 0.02 mag.
The aperture corrections 
from 5 pixels to infinity were taken from the tables of encircled 
energy distributions in \citet{sir05}.

\subsubsection{Completeness and Contamination}

\begin{figure}
\centering
\includegraphics[width=85mm]{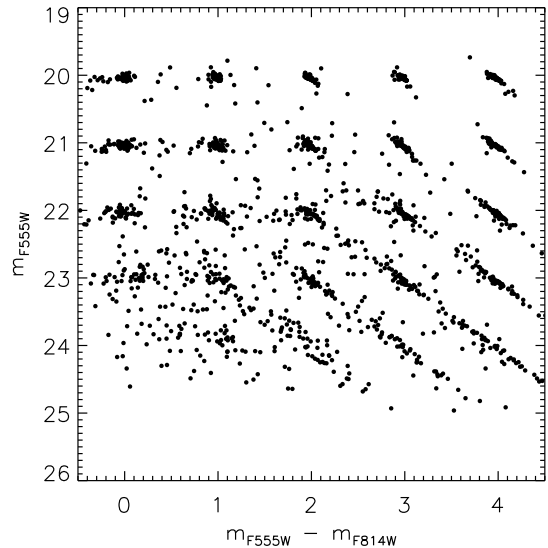}
\caption{\label{fig:scatcmd}
Colour-magnitude diagram for artificial objects after recovery and
PSF-fitting photometry. Objects
with input magnitudes of $m_{\rm F555W} = 20, 21, ... 24$
and $m_{\rm F555W} - m_{\rm F814W} = 0, 1, 2, 3, 4$ are shown here.
}
\end{figure}

\begin{figure}
\centering
\includegraphics[width=85mm]{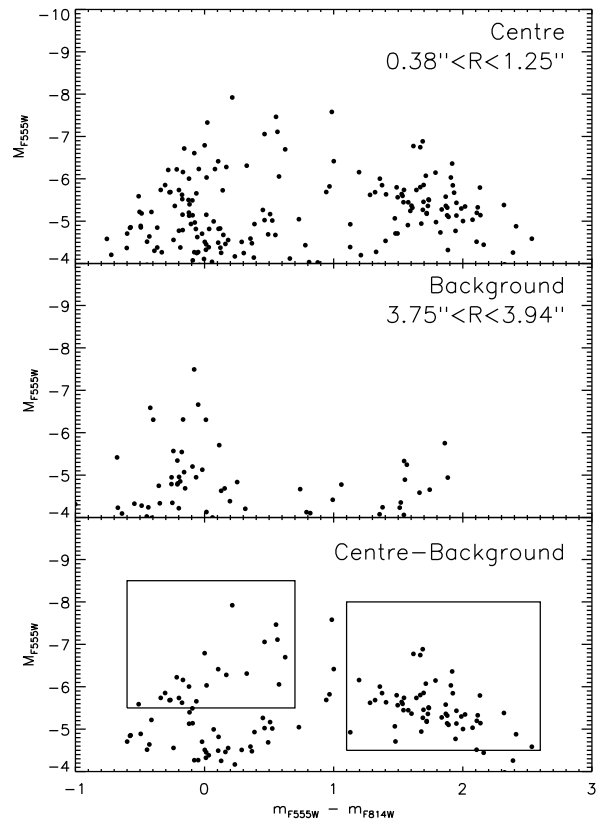}
\caption{\label{fig:cmdVI_sub}
Statistical decontamination of the CMD for NGC~1569-B. Top: the
CMD for the annulus illustrated in Fig~\ref{fig:phot}. 
Centre: the CMD for an annulus of larger
radius, but identical surface area. Bottom: sources in the central panel
which have a match within 0.5 mag in the top panel have been subtracted
from the top panel.
}
\end{figure}

The resulting CMD is shown in Fig.~\ref{fig:cmdraw} for the radial range 
15 pixels $<$ R $<$ 50 pixels ($0\farcs37 < R < 1\farcs25$). Closer
to the centre the crowding becomes too severe to obtain useful
photometry and further out contamination from the general field population
becomes an increasing worry. Our CMD is consistent with the one
already shown by \citet{hunter00}, who used data from the
Wide Field Planetary Camera 2, in containing a
large number of RSGs.  

The severe crowding requires that careful attention be paid to
completeness effects. We quantified the detection incompleteness
and photometric errors by adding artificial stars in the same annulus
for which the CMD was constructed.
A common procedure when carrying out such artificial object tests is to
add stars at random positions within an image; however, it is not 
necessary that the positions are truly random. As long as the
underlying distribution of stars is random itself, no loss of generality
results from adding the artificial stars in a semi-regular grid.
The advantage is that they can be more closely packed this way.  The
only caveat is that care must be taken to sample different positions
of the PSF at the sub-pixel level, in particular for undersampled images such
as those from HST.  We distributed the artificial stars in 
concentric rings with a radial separation of 10 pixels between 
each ring and a separation of 10 pixels between stars within each ring. 
This approach automatically provides the required (pseudo-) randomness of the 
PSF positions at sub-pixel level and ensures that no two artificial stars
are closer together than 10 pixels, thus avoiding ``self-crowding'' of
the artificial stars.

The artificial stars were recovered using exactly the same procedure as
for the regular photometry. In order to sample the incompleteness in
different regions of the CMD, artificial stars with magnitudes
between $m_{\rm F555W} = 20.0$ and $m_{\rm F555W} = 26.0$ 
(in steps of 0.5 mag) and colours between 
$m_{\rm F555W} - m_{\rm F814W} = 0.0$
and
$m_{\rm F555W} - m_{\rm F814W} = 4.0$ (steps of 1 mag) were added. 
By repeating the procedure 5 times, each with a different origin of the
grid, we could sample a total of 346 artificial stars at every magnitude 
and colour within the annulus used for 
the photometry, allowing us to quantify the detection completeness with 
good statistical accuracy.

Fig.~\ref{fig:scatcmd} shows the CMD for the recovered objects. For
clarity, every second magnitude step has been omitted, and only one of
the 5 tests at each colour/magnitude is shown.  A comparison
with Fig.~\ref{fig:cmdraw} shows that the concentration 
of red supergiants at 
$m_{\rm F555W} - m_{\rm F814W} \approx 2.5$
and 
$m_{\rm F555W} \approx 23$ is real and not just an artefact of
incompleteness towards fainter magnitudes. 
If any significant number of red objects with $m_{\rm F555W} \approx 24$ 
were present, they would be clearly detected. On the other hand, 
in the blue part of the CMD the fainter magnitude limit appears to be
predominantly a detection limit. To put these statements on a firmer
ground, we used the output from the completeness tests to calculate the 
magnitudes at which 25\%,
50\% and 75\% of the artificial objects were recovered as a function of
colour. These limits
are shown in Fig.~\ref{fig:cmdraw} and confirm that the concentration
of RSGs is found well above any of these limits. The lower envelope
of the remaining data points is consistent with the completeness
limits inferred from the artificial star experiments.
We conclude that the RSGs, which dominate the
near-IR spectra, are well sampled by the ACS/HRC data, and we can
use the photometry to constrain their properties further.

The artificial star tests also show that much of the scatter in the
colours of the RSGs can be accounted for by photometric errors.
Because the errors on the $m_{\rm F555W}-m_{\rm F814W}$ colours are
dominated by the errors on $m_{\rm F555W}$ for red objects, the
recovered artificial star magnitudes at the position of the RSGs scatter 
along a line which is nearly diagonal in the CMD. A similar trend
is seen in the observed CMD in Fig.~\ref{fig:cmdraw}. This suggests
that the RSGs in NGC~1569-B span a fairly narrow intrinsic range in 
$m_{\rm F555W}-m_{\rm F814W}$ colour, and hence in $T_{\rm eff}$.

Before proceeding, the question of contamination of the cluster CMD
by the general field population of NGC~1569 must be addressed.
Crowding prevents us from measuring stars closer than about $0\farcs38$
from the centre of the cluster, which is about twice the
half-light radius.
In the top panel of Fig.~\ref{fig:cmdVI_sub} we again show the CMD 
within $0\farcs38<R<1\farcs25$, now corrected for Galactic extinction
and converted to an absolute magnitude scale.  The centre panel shows the CMD
for a larger annulus with the same area. Completeness effects
are expected to be less severe in this less crowded region, yet the
population of RSGs is virtually absent. This supports the notion that the 
RSGs in the top panel are indeed predominantly associated with cluster B
as also concluded by \citet{hunter00}.
The bottom panel shows the statistical difference between the top and
centre panels, constructed by removing every star in the central annulus 
which has a matching datapoint within 0.5 mag in $m_{F555W}$ and $m_{F814W}$
in the background annulus.
For the RSGs there is no significant difference, although
some fraction of the blue stars are removed. Since we have not applied
separate completeness corrections in the two annuli, it is likely that 
too many blue stars have been subtracted near the detection limit.

\subsubsection{Comparison with spectroscopy}
\label{sec:speccmp}

Within the right-hand selection box in the bottom panel of 
Fig.~\ref{fig:cmdVI_sub}, we find 57 RSGs with a mean colour of 
$\langle M_{\rm F555W} - M_{\rm F814W}\rangle = 1.75$
and
$\langle M_{\rm F555W}\rangle = -5.6$.
For each star we can estimate the effective temperature and bolometric
corrections from the observed $M_{\rm F555W} - M_{\rm F814W}$ colour,
by using conversions computed with the code described by
\citet{ol00} and multiplying the model atmospheres by
\citet{bes98} with the ACS filter responses.
This yields $\langle T_{\rm eff} \rangle = 3850$ K, very close to the
estimate of $T_{\rm eff} = 3800\pm200$ K obtained from the spectral analysis.
If we instead calculate the $T_{\rm eff}$ values individually for each
star and then compute an average weighted by the estimated $H$-band
luminosity of each star, cooler stars will receive greater weights and
the average $T_{\rm eff}$ decreases to
$\langle T_{\rm eff} \rangle = 3790$ K, again in
excellent agreement with the spectroscopic estimate.

Different assumptions about the distance will not affect the photometric 
estimate of
$T_{\rm eff}$, although different assumptions about extinction clearly will.
If instead of $A_B=2.30$ we assume $A_B=3.02$ \citep{sch98}, then 
$T_{\rm eff}$ increases by about 200 K. 

Having estimated the mean absolute magnitude and effective temperature of the 
RSGs from the CMD, the stellar diameter follows and we can estimate 
the surface gravity if the mass
is known. For an age of $\log t = 7.3$ (\S \ref{sec:interp}), isochrones
from the Padua group \citep{gir00}
give a turn-off mass of 12.4 M$_\odot$. Assuming $T_{\rm eff} = 3800$ K
we get a bolometric correction of $-1.4$ mag from the Kurucz 
models\footnote{Downloaded from http://kurucz.harvard.edu/grids.html}.
Then the mean RSG luminosity is $L_{\rm RSG}/L_\odot = 4.9\times10^4$ and 
the radius is $R_{\rm RSG}/R_\odot = 510$, 
which leads
to $\log g = 0.1$ (in cgs units). This is again very close to the value
$\log g = 0.0$ derived from the RSG model spectral fit.

\subsubsection{Model comparison and the ratio of blue to red supergiants}
\label{sec:interp}

\begin{table}
\begin{center}
\caption{Observed and model supergiant properties. BSG/RSG is the
 ratio of blue to red supergiants and $\langle T_{\rm eff} \rangle_{\rm RSG}$
 is the mean effective temperature of red supergiants, weighted by their
 absolute $M_{\rm F555W}$ magnitude. Red supergiants are here defined as
 stars having $M_{\rm F555W} < -4.5$ and $M_{\rm F555W} - M_{\rm F814W} > 1.1$.
\label{tab:sgprops}}
\begin{tabular}{lcc} \hline
                 &   BSG/RSG     &   $\langle T_{\rm eff} \rangle_{\rm RSG}$ [K]\\ 
		 \hline
Observed         & $0.39\pm0.10$ &   3850          \\
Padua $Z=0.008$  &               &                 \\
\,  log t = 7.0  & 1.75          &   4130          \\
\,  log t = 7.2  & 1.70          &   4110          \\
\,  log t = 7.4  & 0.80          &   4100          \\ 
Geneva $Z=0.008$ &               &                 \\
\,  log t = 7.0  & 0.29          &   3870          \\
\,  log t = 7.2  & 3.29          &   3950          \\
\,  log t = 7.4  & 1.10          &   3970          \\
Padua $Z=0.004$  &               &                 \\
\,  log t = 7.0  & 1.50          &   4350          \\
\,  log t = 7.2  & 2.42          &   4290          \\
\,  log t = 7.4  & 0.81          &   4310          \\ 
Geneva $Z=0.004$ &               &                 \\
\,  log t = 7.0  & 8.69          &   4270          \\
\,  log t = 7.2  & 33.7          &   4110          \\
\,  log t = 7.4  & 2.67          &   4200          \\
\hline
\end{tabular}
\end{center}
\end{table}

\begin{figure*}
\centering
\includegraphics[width=85mm]{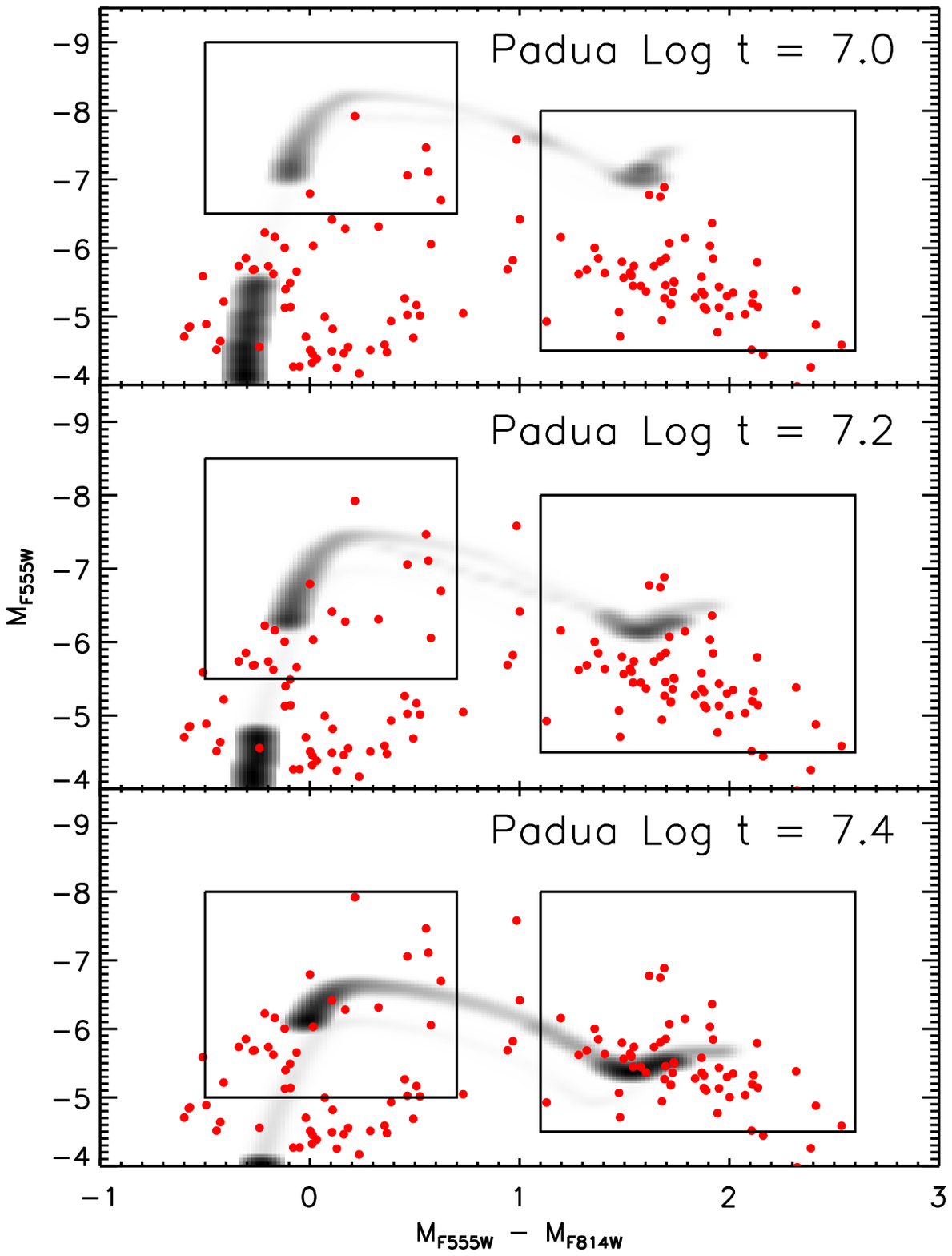}
\includegraphics[width=85mm]{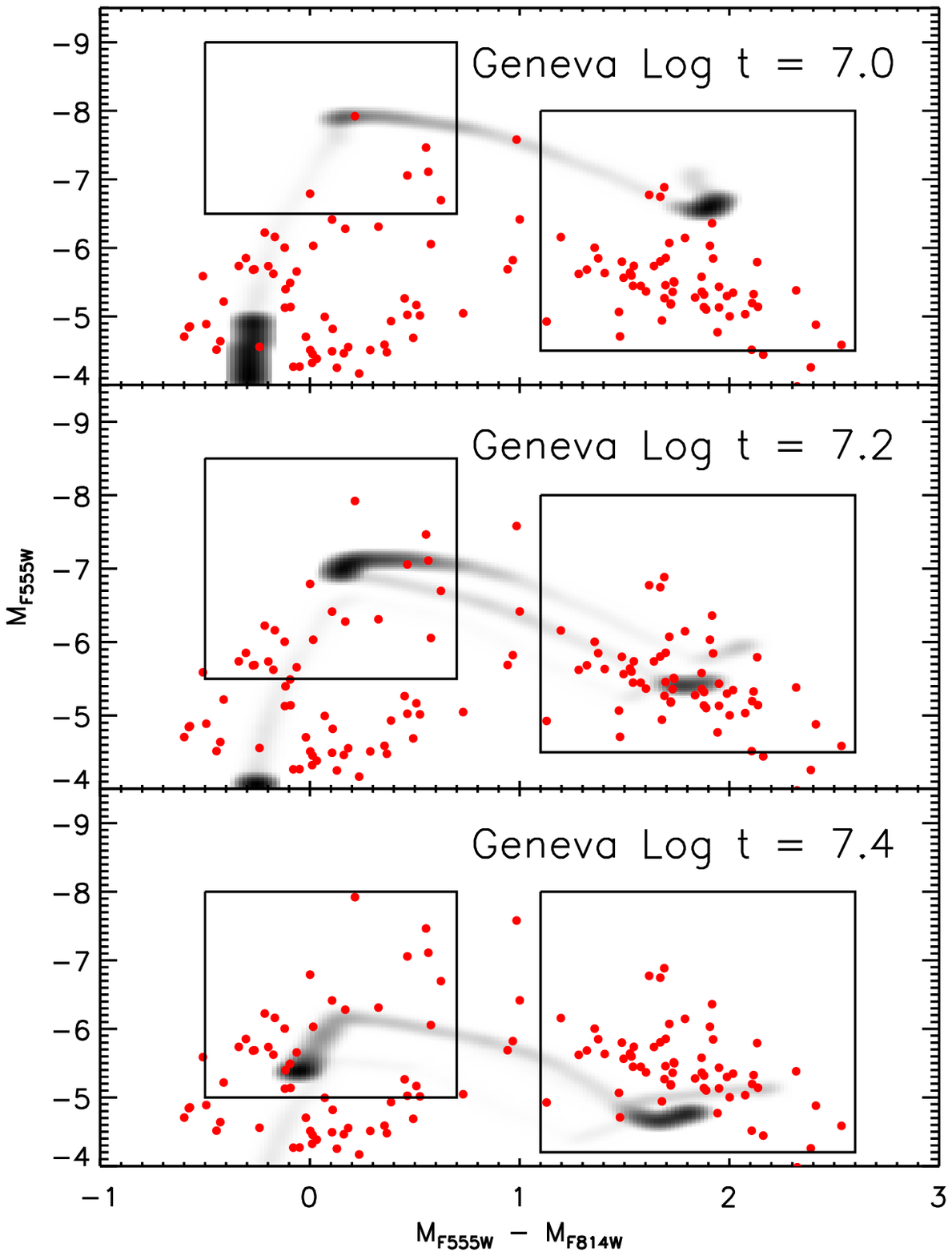}
\caption{\label{fig:hess_pd}
Hess diagrams for Padua (left) and Geneva (right) $Z=0.008$ isochrones 
for three different ages. The filled circles (red in the on-line edition)
are the observed CMD for NGC~1569-B.
}
\end{figure*}

\begin{figure*}
\centering
\includegraphics[width=85mm]{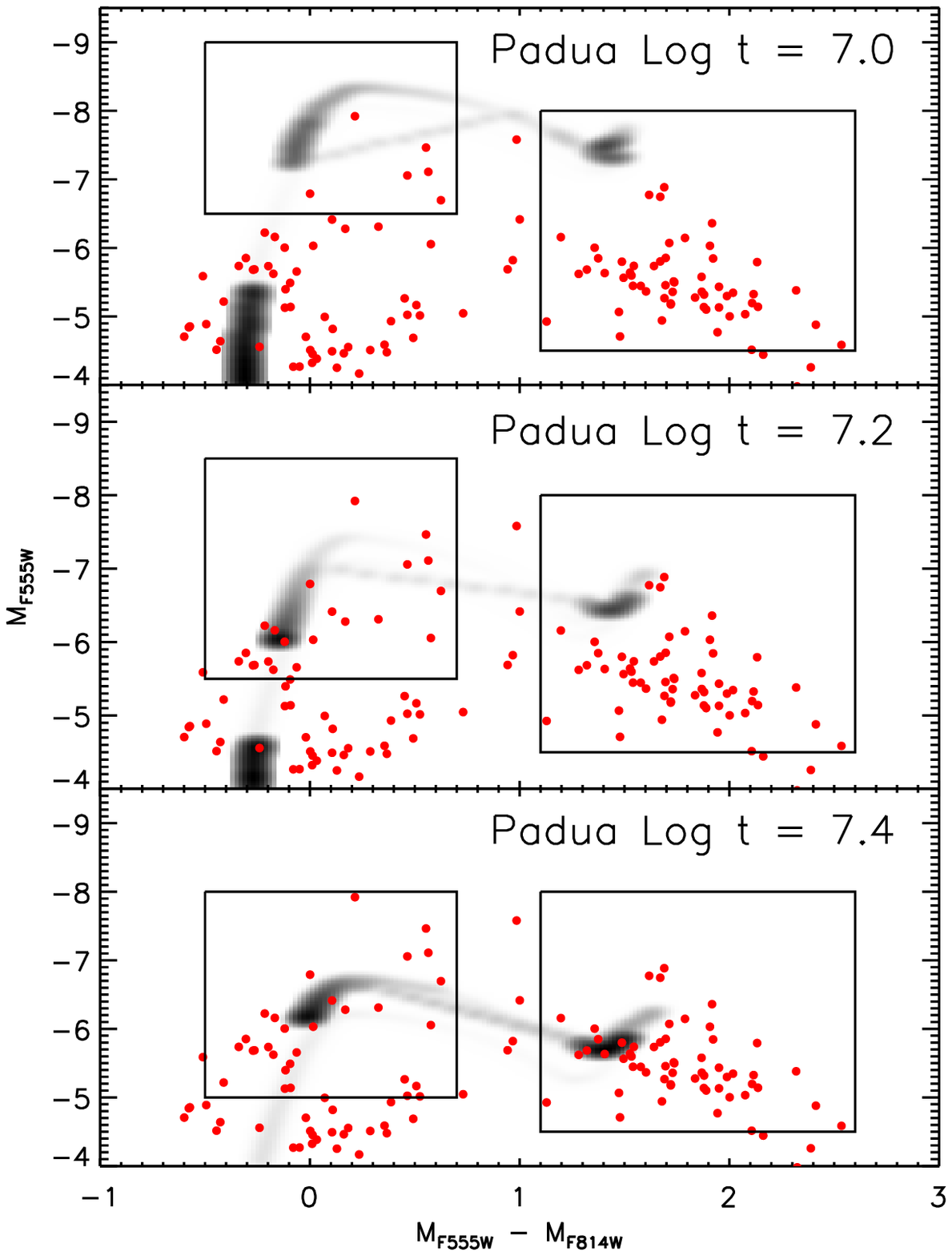}
\includegraphics[width=85mm]{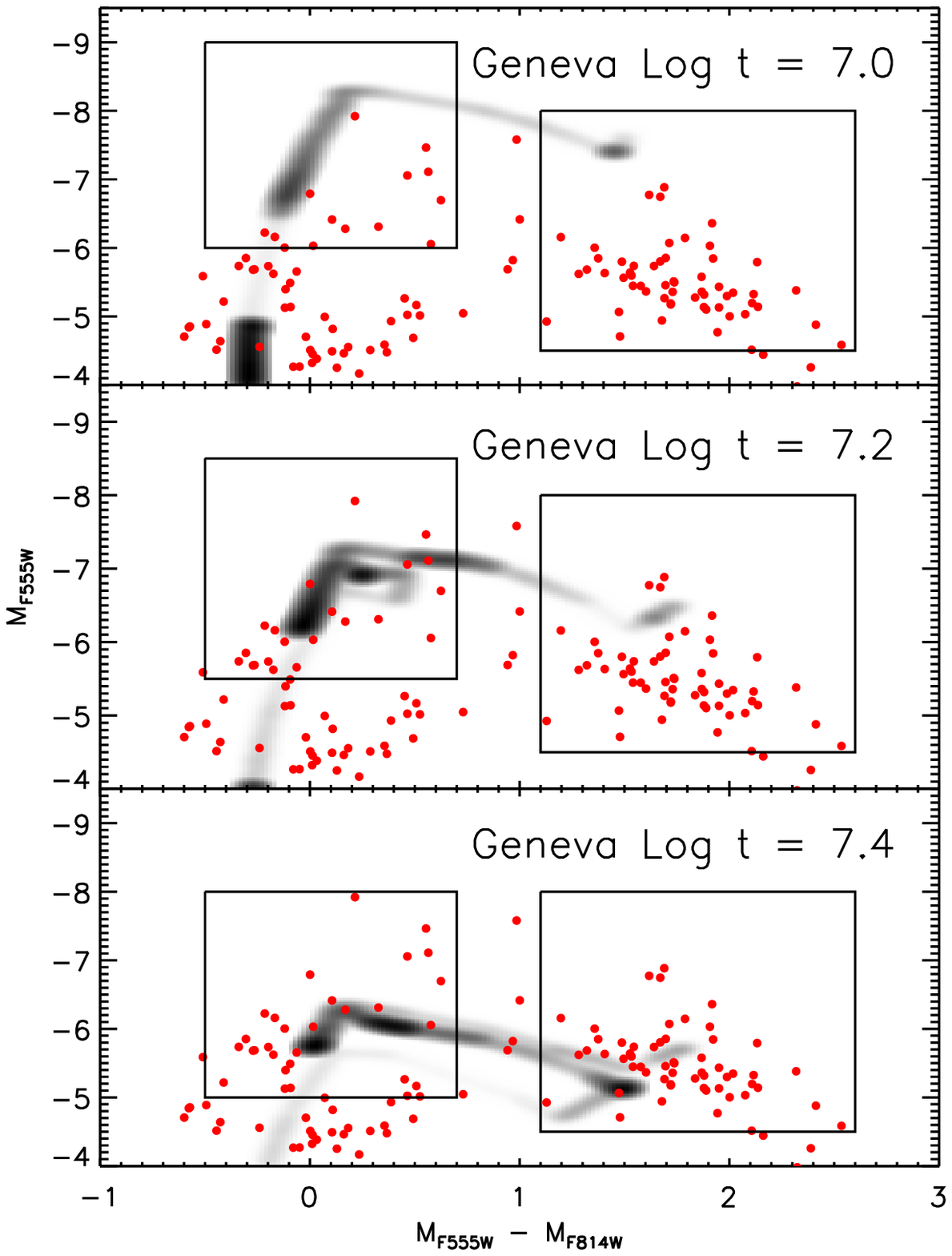}
\caption{\label{fig:hess_pd_z004}
Same as Fig.~\ref{fig:hess_pd} except that isochrones for
$Z=0.004$ are shown.
}
\end{figure*}

Models for massive stars are still fraught with significant uncertainties,
partly because there is still only a limited amount of data available to 
compare with.  CMDs for YMCs such as NGC~1569-B can help constrain the models.

The mean effective temperature of RSGs predicted by models from the
Padua and Geneva \citep{ls01} groups are listed in 
Table~\ref{tab:sgprops} for two different metallicities: $Z=0.004$
and $Z=0.008$. The Geneva models are available for ``standard'' and
``enhanced'' mass loss rates. Here we use the models with enhanced
mass loss since these compare most favourably with the data.
We have selected the RSGs in the
same way as for the empirical data, i.e.\ RSGs are here defined as
stars with $M_{\rm F555W} - M_{\rm F814W} > 1.1$. The 
$\langle T_{\rm eff} \rangle_{\rm RSG}$ values in Table~\ref{tab:sgprops} 
are the average of the $T_{\rm eff}$ values listed in the 
isochrone tables for these stars, weighted by the luminosity in F555W
and the number of stars at each initial mass according to a 
\cite{kroupa02} IMF.
Because all RSGs have very nearly the same mass for a given age, it makes 
virtually no difference what IMF is actually assumed and even for a flat IMF 
the mean effective temperature would change by only a few K. In all cases,
the models predict the RSGs to be hotter than observed, with the
$Z=0.008$ Geneva isochrones coming closest to matching the observations
in this respect. The $Z=0.008$ Geneva models for standard mass loss produce 
RSGs which are about 100 K hotter than the enhanced mass loss models for 
the two youngest ages, while
for the $Z=0.004$ models (which are less affected by mass loss) there
is no difference.

In Fig.~\ref{fig:hess_pd} and \ref{fig:hess_pd_z004} we compare our
background-subtracted CMD with the theoretical isochrones.
Rather than simply drawing the isochrones
as lines connecting the model points, we plot synthetic Hess diagrams 
which show the density of stars predicted by the isochrones. 
The CMDs confirm that the best match to the RSGs is
obtained for $Z=0.008$ with a slightly younger
age being preferred for the Geneva models ($\log t \sim 7.2$) than for the 
Padua models ($\log t \sim 7.4$).  These ages are older than the estimate 
of $\log t = 7.1$ by \citet{and04}, based on integrated broad-band colours. 
However,  if the distance 
of NGC~1569 is greater than we have assumed then the absolute magnitude scale
will shift accordingly, favouring a younger age. 
It seems that the CMD 
favours a metallicity for NGC~1569-B closer to $Z=0.008$ than $Z=0.004$, 
but we are reluctant to stress this point very strongly given the large 
uncertainties in the stellar models for these young ages.

  One specific ``amplifier'' of the uncertainties in models for massive
stars is the
number ratio of blue and red supergiants (BSG/RSG), as discussed in detail by
\citet{lm95}. While it is relatively straight-forward to identify the RSGs in
NGC~1569-B, it is, unfortunately, less clear how to select a sample of 
BSGs for comparison with the models as these are not clearly separated
in the CMD.  The Hess diagrams in Figs.~\ref{fig:hess_pd} 
and \ref{fig:hess_pd_z004} provide some guidance and suggest
that BSGs within the relevant age range should be brighter than
$M_V\approx-5.5$.  We therefore select BSG candidates within the left-hand 
box drawn in the bottom panel of Fig.~\ref{fig:cmdVI_sub}.  This results 
in 22 BSGs and 57 RSGs and thus a ratio of BSG/RSG = $0.39\pm0.10$. 
In Table~\ref{tab:sgprops} we list the predictions by the Padua and
Geneva models, obtained by counting the BSGs and RSGs within the
boxes shown in Fig.~\ref{fig:hess_pd} and \ref{fig:hess_pd_z004}. We have
used the same colour range as in Fig.~\ref{fig:cmdVI_sub}, but adjusted
the magnitude limits for each isochrone to avoid ``contamination'' from 
main sequence stars while making sure that the full range of BSG
luminosities is covered in every case. Only the youngest $Z=0.008$
Geneva isochrone can match the observed low BSG/RSG ratio, and in many
of the other cases the discrepancy is very large,
most notably for the metal-poor Geneva isochrones.

More generous selection limits for BSGs in the observed CMD will reduce the 
discrepancy, but in any case it is clear that the isochrones do not reproduce 
the observed CMD in detail.  The problem of the BSG/RSG ratio will be 
addressed in more detail in forthcoming papers, using a larger sample of 
star clusters of different ages \citep[see also][]{lar07}.

\subsection{Structural Parameters and Dynamical Mass}
\label{sec:virmass}

\begin{figure}
\centering
\includegraphics[width=85mm]{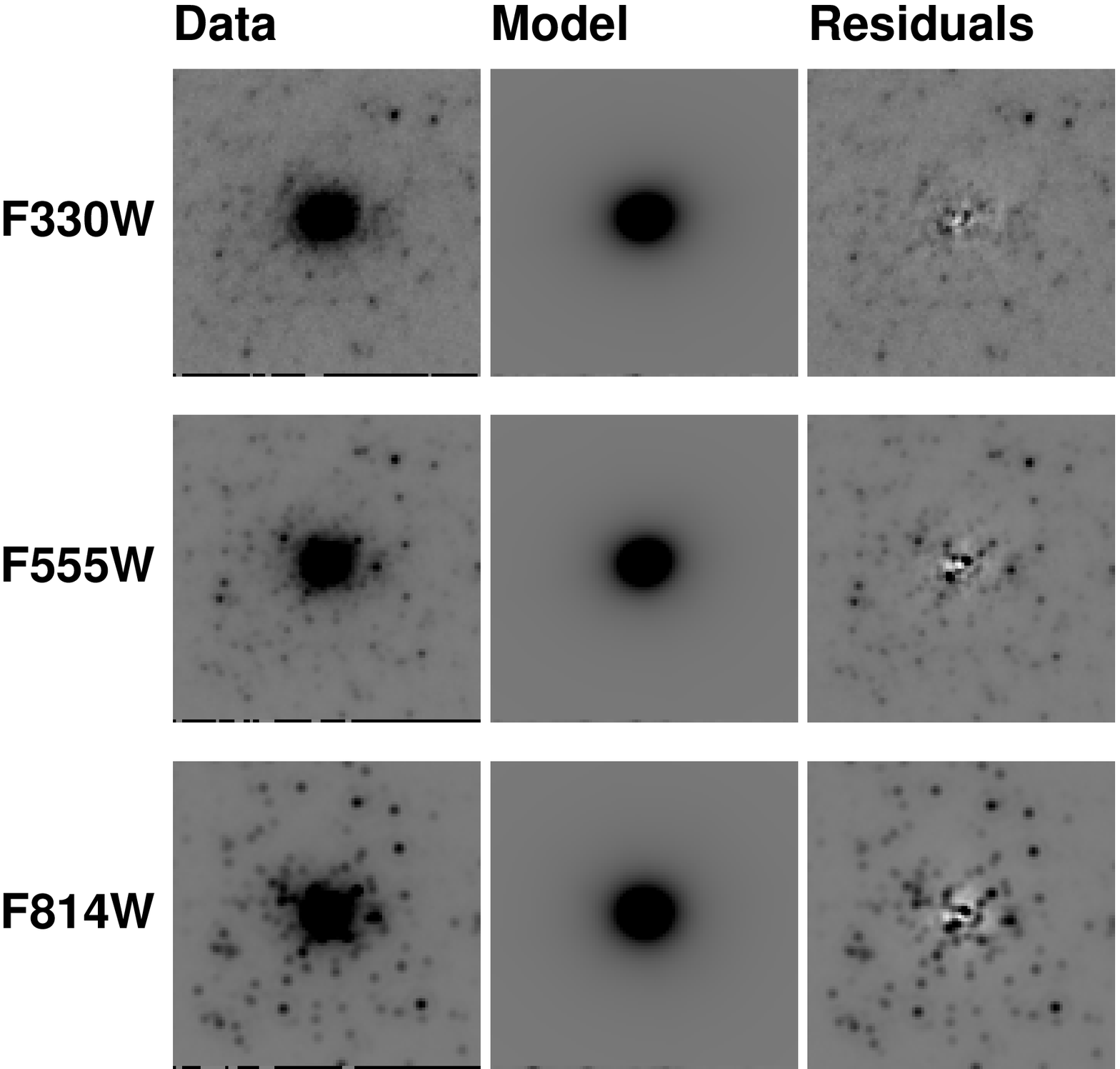}
\caption{\label{fig:fitfig}
HRC images, ISHAPE model fits and residuals of NGC~1569-B in three passbands. 
These panels show a $100\times100$ pixels section, corresponding to
$2\farcs5\times2\farcs5$.
}
\end{figure}

\begin{table}
\begin{center}
\caption{Structural parameters for NGC~1569-B. The FWHM and $R_h$ are
given in pixels (1 pixel = $0\farcs025$ = 0.27 pc (D / 2.2 Mpc)).
\label{tab:struct}}
\begin{tabular}{lccccc}
\hline
   & FWHM & B/A  & $\eta$  &  $R_h(\infty)$ & $R_h(100)$ \\
   & [pixels] &  &         &  [pixels]      & [pixels] \\
\hline
\multicolumn{3}{l}{Fitrad=15 pixels} \\
F330W &  4.72 & 0.77 & 1.16 & 20.02 &  8.32 \\
F555W &  4.27 & 0.82 & 1.30 &  6.98 &  5.72 \\
F814W &  3.68 & 0.86 & 1.32 &  5.72 &  4.92 \\
\multicolumn{3}{l}{Fitrad=20 pixels} \\
F330W &  4.76 & 0.77 & 1.17 & 17.84 &  8.16 \\
F555W &  3.23 & 0.81 & 1.02 & $5\times10^7$ & 10.54 \\
F814W &  3.14 & 0.81 & 1.03 & $1.5\times10^5$ &  9.98 \\
\multicolumn{3}{l}{Fitrad=25 pixels} \\
F330W &  5.33 & 0.75 & 1.31 &  8.07 &  6.56 \\
F555W &  4.54 & 0.79 & 1.32 &  6.79 &  5.72 \\
F814W &  3.68 & 0.81 & 1.18 & 12.64 &  6.66 \\
\multicolumn{3}{l}{Fitrad=35 pixels} \\
F330W &  5.11 & 0.75 & 1.26 &  9.55 &  6.96 \\
F555W &  4.24 & 0.80 & 1.27 &  7.77 &  5.98 \\
F814W &  3.83 & 0.82 & 1.20 & 10.97 &  6.55 \\
\multicolumn{3}{l}{Fitrad=50 pixels} \\
F330W &  4.78 & 0.75 & 1.14 & 27.08 &  8.76 \\
F555W &  3.81 & 0.79 & 1.16 & 16.34 &  7.17 \\
F814W &  3.19 & 0.86 & 1.10 & 50.64 &  7.95 \\
\hline
\end{tabular}
\end{center}
\end{table}

\begin{table}
\begin{center}
\caption{A summary of properties of NGC~1569-B. $R_h$ is the half-light
radius, $m_{\rm F555W}$ is our estimated total apparent magnitude in
the F555W band, $\sigma_r$ is the projected velocity dispersion, 
$v_r$ the heliocentric radial velocity, $M_{\rm vir}$ the dynamical
mass and $\rho_h$ the mean density within the half-mass radius.
For quantities that depend on the assumed distance this dependency
is specified explicitly.
\label{tab:props}}
\begin{tabular}{lc}
\hline
$R_h$ (arcsec)                & $0\farcs20\pm0\farcs05$ \\
$m_{\rm F555W}$ (VEGAMAG)     & $15.60\pm0.20$ \\
$\sigma_r$ (km/s)             & $9.6\pm0.3$ \\
$v_r$ (km/s)                  & $-78\pm3$ \\
$R_h$ (pc)                    & $(2.1\pm0.5) \, (D/2.2\,{\rm Mpc})$  \\
$M_{\rm vir}$ ($M_\odot$)     & $(4.4\pm1.1) \, 10^5 \, (D/2.2\,{\rm Mpc})$ \\
$\rho_h$ ($M_\odot$ pc$^{-3}$) & $(2.5\pm1.3) \, 10^3 \, (D/2.2\,{\rm Mpc})^{-2}$ \\
\hline
\end{tabular}
\end{center}
\end{table}

\subsubsection{Cluster size and its Wavelength Dependence}
\label{sec:size}

Having discussed the stellar content of NGC~1569-B, we
now turn to a reexamination of its structural parameters and
dynamical mass. 
We measured the size
of the cluster in the ACS/HRC images using our ISHAPE
code \citep{lar99}. The cluster size is determined by convolving a
series of analytic model profiles with the PSF and varying the 
model parameters until the best fit is obtained. We used ``EFF'' 
profiles \citep{eff87} of the form
\begin{equation}
  I(R) = I_0 \left[ 1 + (r/r_c)^2 \right]^{-\eta},
  \label{eq:eff}
\end{equation}
where $r_c$ is the core radius and $\eta$ determines the steepness of the
profile at large radii.
Because neither the cluster itself nor the fitting profile has
a clearly defined outer boundary, we experimented with various
fitting radii up to 50 pixels (15 pc).  Fig.~\ref{fig:fitfig} shows
the image of the cluster, the best-fitting model, and the 
residuals for each band, for the largest fitting radius of 50 pixels.  The
resolution into individual stars means that the $\chi^2_\nu$ values of
the fits never get close to unity, but otherwise the models provide a 
good fit to the data.

The PSFs used for the size measurements were based on DAOPHOT PSFs
in the inner regions ($R<5$ pixels).  We know from the smooth residual
images produced by the ALLFRAME photometry that these PSFs accurately
represent the PSFs of our images.  However, for the ISHAPE analysis we
need to know the PSF at larger radii than for the photometry, and 
there the construction
of accurate empirical PSFs becomes limited by decreasing S/N and
non-uniformity of the background.  For the analysis of structural parameters
we therefore made use of composite PSFs, constructed from the DAOPHOT
PSFs for $R<5$ pixels and using models computed by the TINYTIM package
\citep{kh97} further out.

Table~\ref{tab:struct} lists the Full Width at Half Maximum (FWHM) along
the major axis, the minor/major axis ratio (B/A), the slope parameter
$\eta$, and the half-light radius $R_h$ for various fitting radii. 
All fits consistently yield an axis ratio of B/A$\sim0.8$, confirming the 
conclusion from earlier work that NGC~1569-B is significantly elongated
\citep{oco94}.  
For $\eta\leq1$ the total luminosity of the model profile is infinite,
and $R_h$ therefore undefined.  For some fits $\eta$ is close 
to unity, leading to unreasonably large $R_h$ values. We 
list two values for $R_h$: One assuming that the profile extends to
infinity ($R_h(\infty)$) and another assuming that it is
truncated at $R=100$ pixels ($R_h(100)$). For $\eta \gg 1$ the two
$R_h$ values are similar, but for $\eta\approx1$ they differ substantially.
The cut at $R=100$ pixels (30 pc) is chosen fairly arbitrarily, apart
from the fact that it is very difficult to trace the cluster profile
to greater radii in the images.

\begin{figure}
\centering
\includegraphics[width=85mm]{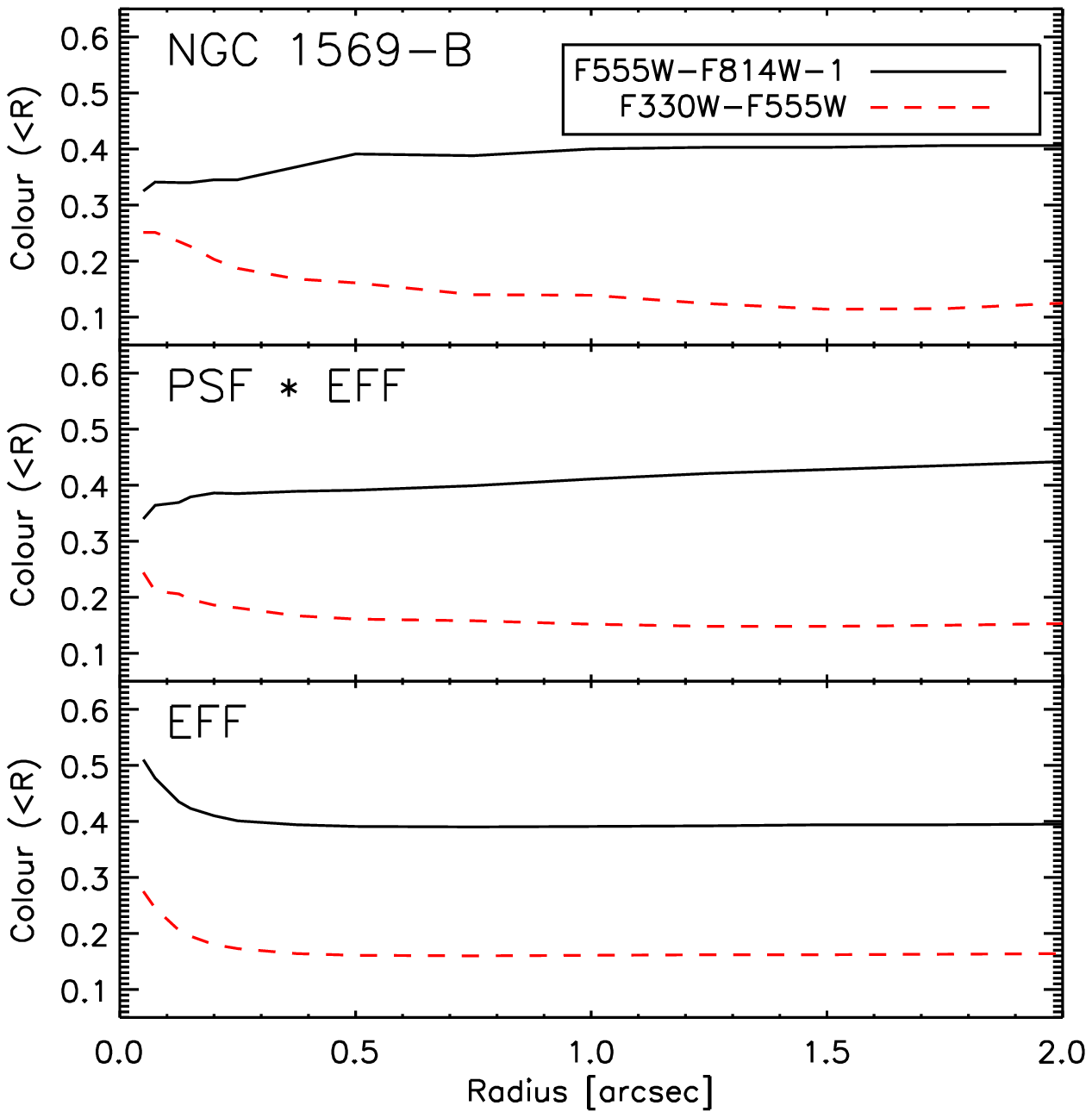}
\caption{\label{fig:colprof}Top: $m_{\rm F330W}-m_{\rm F555W}$ and
$m_{\rm F555W}-m_{\rm F814W}$ colour versus aperture size for
NGC~1569-B. To fit the two plots on the same scale, 1 magnitude has
been subtracted from $m_{\rm F555W}-m_{\rm F814W}$.
Centre: Same, for the PSF in each filter convolved with EFF profiles
of FWHM = 5.0, 4.5 and 4.0 pixels in F330W, F555W and F814W.
The model profiles have been normalised to the same
colour as the observed profiles at a radius of $0\farcs5$.
Bottom: Same, but for the EFF profiles without any PSF convolution.
}
\end{figure}

Regardless of the fitting radius, all fits return a smaller FWHM 
at longer wavelengths. The formal errors on the FWHM values returned by 
ISHAPE are generally less than $\pm0.1$ pixels, so the trend with wavelength 
appears significant.  
The FWHM is roughly equal to twice the core radius $r_c$ of the EFF
profiles (for $\eta = 1.0$
this holds exactly) so the trend in FWHM with wavelength is basically
a trend in core radius.
Such a trend seems consistent with the finding by \citet{hunter00} that 
NGC~1569-B has redder $m_{\rm F336W}-m_{\rm F555W}$ colours near the centre. 
However, they also found that the $m_{\rm F555W}-m_{\rm F814W}$ colours 
become \emph{bluer} near the centre. In the top panel of Fig.~\ref{fig:colprof} 
we show the $m_{\rm F330W}-m_{\rm F555W}$ and $m_{\rm F555W}-m_{\rm F814W}$
colours versus aperture radius for NGC~1569-B measured on the HRC data. The 
trends seen in this
plot are very similar to those observed by Hunter et al., with 
$m_{\rm F330W}-m_{\rm F555W}$ becoming redder inwards and 
$m_{\rm F555W}-m_{\rm F814W}$ getting bluer. How can this be consistent
with a monotonic decrease in cluster core radius with wavelength?

The answer appears to be that a na{\"i}ve comparison of colour gradients,
as in Fig.~\ref{fig:colprof}, neglects the effect of the PSF.
Colour gradients can result
from a combination of changes with wavelength in the structure of 
\emph{both} the cluster \emph{and} the PSF.
The PSF behaviour with wavelength is not monotonic.
Due to the ``red halo''
effect, more light is scattered to large
radii at longer wavelengths, but the encircled energy distributions 
tabulated in \citet{sir05} also show that more light is scattered to 
intermediate radii ($0\farcs1$ -- $1\arcsec$) in F330W relative to F555W.
This is also illustrated by the fact that the aperture corrections
to infinity display a minimum around 5000\AA -- 6000\AA\ 
\citep[Fig.\ 10 in][]{sir05}. 

To investigate this further, we convolved the PSF in F330W, F555W and F814W 
with EFF models for the respective filters having FWHM of 5.0, 4.5 and 4.0 
pixels and envelope slopes 
of $\eta = 1.26$, 1.23 and 1.20 (guided by the numbers in 
Table~\ref{tab:struct}).  The centre panel of Fig.~\ref{fig:colprof} shows 
the integrated colours versus aperture radius for these model profiles, 
normalised to match the observed colours at $0\farcs5$. Although the intrinsic 
cluster core radius decreases monotonically with wavelength by design,
we reproduce the observed trend towards bluer
$m_{\rm F555W}-m_{\rm F814W}$ colours and redder
$m_{\rm F330W}-m_{\rm F555W}$ colours near the centre. The bottom panel
shows the colour-aperture relations for the pure EFF models, as they would be
observed in the absence of any convolution with the PSF. In this case
both the $m_{\rm F555W}-m_{\rm F814W}$ and $m_{\rm F330W}-m_{\rm F555W}$
colours become redder near the centre.
We conclude that the colour gradients observed by \citet{hunter00} (and
also seen in the HRC data) are \emph{consistent} with the monotonic decrease 
in the cluster FWHM with wavelength obtained from the ISHAPE fits.

Because the envelope slope $\eta$ tends to be shallower at longer 
wavelengths, the correlation between FWHM and wavelength does not 
translate into a similarly strong trend of half-light radius $R_h$ versus 
wavelength.  The $R_h(\infty)$ values show a very larger scatter
while most $R_h(100)$ values are in the range 6--8 pixels or
$0\farcs15$--$0\farcs20$. \citet{oco94} measured a half-light radius
of $R_h = 0\farcs25$ on HST/WFPC images, while \citet{hunter00} derived 
an almost identical value of $R_h = 0\farcs26$ from HST/WFPC2 images. 
Our measurement on the ACS/HRC images is slightly smaller, but
an increase in the cut-off radius by a factor of 2--3 would bring our 
$R_h$ estimate into close agreement with previous studies.
In the following, we therefore adopt $R_h=0\farcs20\pm0\farcs05$.

\subsubsection{Mass segregation?}

NGC~1569-B is not the only YMC in which the structure has been found to
be wavelength dependent.
\citet{laretal01} noted that the colours of a YMC in NGC~6946
become bluer in larger apertures.
A trend of smaller cluster size at longer wavelengths was also found for 
the cluster M82-F by \citet{mgv05}, who argued that it might be due to mass 
segregation 
(although this result has been questioned by \citet{bas07} after
a detailed analysis of the extinction near M82-F).
Because the RSGs are the most massive stars currently
in the cluster, mass segregation would cause them to be more centrally
concentrated, making the cluster appear smaller at longer wavelengths
where the RSGs dominate. 

Mass segregation may be either primordial
in nature (the most massive stars forming preferentially near the centre)
or a result of dynamical evolution; in the latter case it is
expected to occur on the half-mass relaxation time-scale which is
\begin{equation}
  t_{rh} = \frac{1.7\times10^5 [R_h({\rm pc})]^{3/2} N^{1/2}}{[m/M_\odot]^{1/2}} \, {\rm years}
  \label{eq:trh}
\end{equation}
where $N$ is the number of stars in the cluster and $m$ is the mean
mass per star \citep{spit87}. For a total cluster mass of 
$\sim4\times10^5$ M$_\odot$ (\S \ref{sec:dynmass}), a mean stellar mass of 
0.6 M$_\odot$ (for a Kroupa IMF extending from 0.1 M$_\odot$ to 100
M$_\odot$) and $R_h \approx 2$ pc, Eq.~(\ref{eq:trh}) yields 
$t_{rh} \approx 5\times10^8$ years,
about a factor of 25 longer than the current age of NGC~1569-B. Thus,
the cluster is not expected to be fully mass segregated, but this
statement is subject to a few caveats. Near the centre of the cluster,
where the density is higher, the relaxation time will be shorter.
Furthermore, massive stars can ``sink'' to the centre on shorter time scales 
due to energy equipartition. The timescale for this to occur is inversely 
proportional to the mass of the stars in question \citep{ger00}, so stars 
with masses characteristic of the current RSGs might become segregated on time 
scales similar to the current age of NGC~1569-B.  According to \citet{mcm07},
dynamical mass segregation may occur even more rapidly if a cluster forms
from subclumps which already have some degree of primordial mass segregation.
Alternatively, mass segregation may be imprinted in the structure of
the cluster from the beginning \citep[e.g.\ references in][]{mcm07}.

Whether or not mass segregation can actually explain the observed relations 
between cluster structure and wavelength is another question.
\citet{fleck06} calculated the
difference $\Delta_{V-K}$ between the $V-K$ colour inside and outside the 
half-light radius for a
mass-segregated cluster as a function of time and found this to be
0.05 -- 0.15 mag. Using simple stellar population models by \citet{bc03}
we estimate $\Delta_{V-K}/\Delta_{V-I} \approx 2$, so we may expect mass
segregation to affect $\Delta_{V-I}$ by a few times 0.01 mag.  From the
data plotted in the bottom panel of Fig.~\ref{fig:colprof} we measure
a difference $\Delta_{V-I} = 0.03$ mag, which appears roughly consistent
with the calculations by Fleck et al. So it seems that mass segregation
\emph{might} be responsible for the observed colour gradients,
although it would be desirable to investigate this in more detail.

\subsubsection{Integrated magnitude}

\begin{figure}
\centering
\includegraphics[width=85mm]{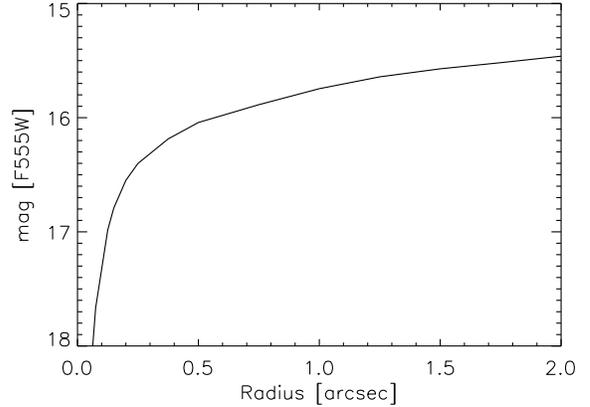}
\caption{\label{fig:cog}
Integrated magnitude versus aperture radius for NGC~1569-B.
The background was measured in an annulus between $2\farcs0$
and $3\farcs0$.
}
\end{figure}

For many of the same reasons that complicate measurements of the 
core- and half-light radius, the integrated magnitude of NGC~1569-B remains 
uncertain and a range of numbers can be found in the literature.
\citet{and04} list $M_{\rm F555W} = -12.83$ for an $R=0\farcs46$
aperture.  For their assumed distance of 2.2 Mpc, combined with reddening 
from Schlegel et al. and a correction for internal reddening in NGC~1569 of 
$E(B-V) = 0.05$ mag, this corresponds to $m_{\rm F555W} = 16.30$. 
\citet{origlia2001} list a reddening corrected apparent magnitude of
$m_{\rm F555W,0} = 14.5$ or $m_{\rm F555W} = 16.24$, close to the value
given by Anders et al.  Using a larger
aperture of $R=1\farcs34$, \citet{hunter00} list $M_{\rm F555W} = -13.02$ 
for $E(B-V) = 0.56$ and a distance of $D=2.5$ Mpc, i.e.\
$m_{\rm F555W} = 15.70$. 

Most of the differences between published photometry can be ascribed to
different choices of aperture radii.  Using the same aperture radii, we 
can largely reproduce the numbers given above. For the apertures and
background annuli used by
\citet{and04} and \citet{hunter00}, we obtain 
$m_{\rm F555W}(0\farcs46) = 16.20$ mag 
and $m_{\rm F555W}(1\farcs34) = 15.66$ mag. 
Figure~\ref{fig:cog} shows the integrated apparent 
$m_{\rm F555W}$ magnitude
as a function of aperture radius, demonstrating that the
curve-of-growth never levels off completely out to at least $2\arcsec$. This 
is a generic difficulty for many young star clusters, which tend to
have very extended outer envelopes with $\eta$-values close to 1 and
hence poorly defined total luminosities \citep{lar04}. 

Per definition, the total luminosity of the cluster should be twice that
contained within the effective radius.  We can therefore estimate the
total integrated magnitude by measuring the magnitude within
$R_h$ and doubling the flux.  From the data in Fig.~\ref{fig:cog} we get
$m_{\rm F555W} = 16.79 / 16.55 / 16.40$ mag in apertures of
$0\farcs15/0\farcs20/0\farcs25$. Doubling the fluxes we then
get $m_{\rm F555W}(\infty) \approx 15.80\pm0.20$, where the quoted
error is due to the uncertainty on the half-light radius.  This is 
consistent with the measurement of \citet{hunter00} but brighter than 
measurements made in smaller apertures. However,
this estimate of the integrated magnitude
may still be a lower limit, because we have used a
half-light radius that was already corrected for broadening by the PSF.
In reality, some fraction of the cluster light is scattered from the
centre to larger radii by the PSF, and
we therefore expect that an aperture with the same radius as the intrinsic
$R_h$ will contain slightly \emph{less} than half of the total
cluster luminosity.
By carrying out aperture photometry on a simulated PSF, we estimate this 
effect to be about 0.20 mag, consistent with the encircled energy tables in
\citet{sir05}. Our final estimate of the integrated cluster magnitude
is then $m_{\rm F555W}(\infty) \approx 15.60\pm0.20$.

\subsubsection{Velocity dispersion}
\label{sec:vd}

\begin{figure}
\centering
\includegraphics[width=85mm]{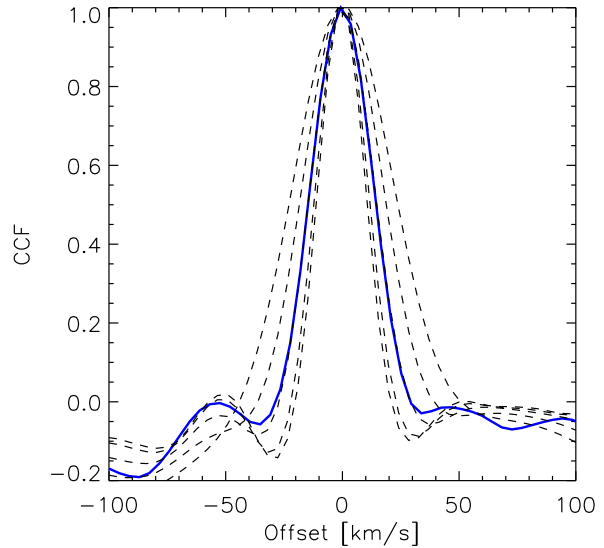}
\caption{\label{fig:ccf}
Cross-correlation functions for order 48 of the NIRSPEC data.
The CCF for NGC~1569-B versus
HR2184 is shown as a thick line (blue in the on-line edition). Thin dashed 
lines show the CCF for HR2235 versus HR2184, with the spectrum of HR2235 being 
broadened by 1 km/s, 5 km/s, 10 km/s, 15 km/s and 20 km/s. The narrowest CCF 
peaks correspond to the smallest amount of broadening. All CCFs have been 
shifted to a mean of 0 km/s for easier comparison.
}
\end{figure}

\begin{figure}
\centering
\includegraphics[width=85mm]{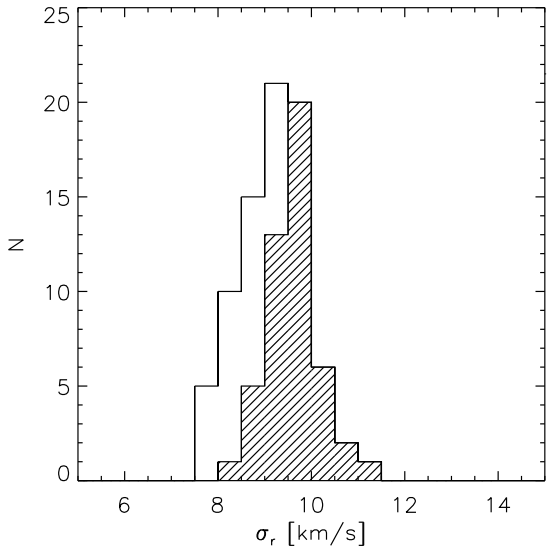}
\caption{\label{fig:vdhist}
Distribution of velocity dispersion measurements from the
cross-correlation analysis. The filled histogram shows the measurements
based on the best orders (45, 46, 49) while the outlined histogram also
includes orders 47 and 48. The best orders yield a mean
of $\sigma_r = 9.6\pm0.1$ km/s.
}
\end{figure}

As noted in \S\ref{sec:abundance}, the best fit of our RSG model
spectrum to the NIRSPEC data requires a further broadening of $9.9\pm1$ km/s
after accounting for the instrumental resolution. 
We have also derived the velocity dispersion using the standard
cross-correlation technique \citep{td79,hf96} applied in many previous studies.
We refer to our previous papers on this subject for details
\citep[e.g.][]{laretal01,lr04,lbh04}. We used two template stars, 
HR~2184 (spectral 
class K1II) and HR~2235 (F5 II).  Fig.~\ref{fig:ccf} shows the 
cross-correlation
function (CCF) for NGC~1569-B versus HR 2184 (thick curve) together with 
the CCF for the spectrum of HR 2235 (broadened by various amounts) versus 
that of HR~2184. The
best match is obtained for a broadening of $\approx10$ km/s.

In Fig.~\ref{fig:vdhist} we show the distribution of velocity broadenings
derived from the cross-correlation analysis in the various orders of
the NIRSPEC spectrum and
the different combinations of template stars. The best 
fits (highest
CCF peaks) were obtained for orders 45, 46 and 49 and the results for these
orders are shown as a filled histogram. The mean is $\sigma_r = 9.6\pm0.1$
km/s. If we include orders 47 and 48 the mean decreases to
$\sigma_r = 9.2\pm0.1$ km/s, still significantly greater than the 
7.5 km/s measured by \citet{gg02}. For the following discussion we will
adopt $\sigma_r = 9.6\pm0.3$ km/s, including the estimate of $\sigma_r$
from the RSG model fit within our 1$\sigma$ error.

\subsubsection{Dynamical mass}
\label{sec:dynmass}

Assuming virial equilibrium and that mass traces light, the cluster
mass is related to the 1-D velocity dispersion $\sigma_r$ and 
half-light radius by
\begin{equation}
  M_{\rm vir} = \alpha \frac{\sigma_r^2 R_h}{G}
  \label{eq:vir}
\end{equation}
with $\alpha\approx9.75$. Inserting the $R_h$ and $\sigma_r$ values
derived in \S \ref{sec:size} and \S \ref{sec:vd} 
(summarised in Table~\ref{tab:props}), we obtain a dynamical mass of
$M_{\rm vir} = (4.4\pm1.1)\times10^5 M_\odot$ which is about
a factor of two higher than the value of $2.3\times10^5$ M$_\odot$
derived by \citet{gg02}. The difference is mostly due to our greater
velocity dispersion. 
Note that
$M_{\rm vir}$ scales linearly with $R_h$, and thus with the assumed
distance $D$. Rather
than trying to include an uncertainty on the distance, we have specified
this dependency explicitly for the cluster mass listed in 
Table~\ref{tab:props}. This dynamical
mass of NGC~1569-B is much greater than for any young star cluster 
currently known in the Milky Way and also well above the median mass
of old globular clusters (though still comfortably within the mass
range of Milky Way GCs).

We can compare the dynamical mass estimate with the mass expected
for a cluster of this age and luminosity, given a normal stellar IMF.
For $Z=0.008$ and a Chabrier IMF, the \citet{bc03} SSP models predict
a $V$-band absolute magnitude of $M_V = 1.53\pm0.15$ mag per $M_\odot$ for 
$\log t = 7.3\pm0.1$.  For $M_V \approx m_{\rm F555W} - 26.7 - A_V = 
-12.85\pm0.2$ the ``photometric'' mass is then $(5.5\pm1.3)\times10^5$ 
$M_\odot$, which is in agreement with the dynamical mass within the 
1 $\sigma$ errors.

The close agreement between the dynamical and photometric mass estimates
is probably somewhat fortuitous, given the uncertainties on the distance, 
reddening and age.  In the absence of reddening, the dynamical-to-photometric 
mass ratio will scale as $1/D$. However, since the problem of determining the 
reddening and distance to NGC~1569-B are closely tied, the comparison is not 
as straight forward. To first order, the absolute magnitude of
NGC~1569-B may be roughly independent of the reddening correction, since 
most distance determinations to NGC~1569 are based on standard candles. 
The dynamical-to-photometric mass ratio will then scale as $D$ instead 
of $1/D$. All this neglects any possible dependency of the age
estimate on distance and reddening. This will depend on the method used
for age determination - ages based on integrated colours (or
spectroscopy) will be distance independent and probably only weakly
dependent on the reddening (since this is usually included as a free
parameter in the fit) while CMD-based ages will depend on both. 

Another source of uncertainty is the parameter $\alpha$ in Eq.~(\ref{eq:vir}).
In a mass segregated cluster, $\alpha$ will have a different value
than we have assumed. Because of
energy equipartition, the massive stars will acquire lower velocities
and sink to the centre. Since \emph{measurements} of the velocity dispersion
and cluster size based on integrated light are dominated by these stars, 
the mass derived from Eq.~(\ref{eq:vir}) will be underestimated.
\citet{fleck06} model the evolution of $\alpha$ for dynamically 
mass segregated star clusters and give an analytical approximation to the
relative error $\Delta \alpha/\alpha$ as a function of age and 
relaxation time.  For an age of 20 Myr and $t_{rh} = 500$ Myrs, their
equation (31) yields $\Delta\alpha/\alpha\approx13$\%. With this
correction, the dynamical mass becomes 
$M_{\rm vir,corr} = (4.9\pm1.2)\times10^5$ M$_\odot$ for $D=2.2$ Mpc,
i.e.\ even in even closer agreement with the photometric estimate.

\section{Discussion and Summary}
\label{sec:summary}

We have presented a detailed investigation of the stellar content and
other properties of the massive star cluster 'B' in NGC~1569. Using
new high S/N $H$-band echelle spectroscopy from the NIRSPEC spectrograph
on Keck~II, we have carried out abundance analysis of red supergiant
stars in the cluster. We find an iron abundance of [Fe/H] = $-0.63\pm0.08$,
close to that of SMC field stars. Our estimate of the oxygen abundance,
[O/H] = $-0.29\pm 0.07$, is about a factor of two higher than that derived
for H{\sc ii} regions \citep{sdk94,dev97,ks97}, and the resulting 
super-solar [O/Fe] abundance is unlike that observed in the Magellanic Clouds
and young stellar populations in the Milky Way. However, according to
our measurements NGC~1569-B also differs from these galaxies by having
a higher [$\alpha$/Fe] ratio.  The difference between our
[O/H] measurement for NGC~1569-B and the O abundance derived for H{\sc ii} 
regions is greater than the formal uncertainties on either value, and our
oxygen abundance is higher than the nebular abundances in most dwarf
galaxies \citep[e.g.][]{vad07}. This
raises the question whether one (or both) methods suffers from possible
systematic errors, or if there might be a genuine difference between 
cluster and H{\sc ii} region abundances in NGC~1569 -- perhaps due to an
ab initio oxygen enhancement of the gas out of which the cluster formed.
Since oxygen is the
only element in common between the cluster and H{\sc ii} region data, a 
detailed comparison of the
abundance patterns is unfortunately not possible. However, it is of interest
to note that the very strong Wolf-Rayet features in the spectrum of
cluster A also suggest a high metallicity \citep{maoz01}.

NGC~1569 is a popular target for testing models of chemical evolution, and
it is generally recognised that the strong galactic wind observed in the
galaxy must play an important role. \citet{martin02} observed a 
super-solar $[\alpha/{\rm Fe}]$ ratio for the wind, as we do for NGC~1569-B.
However, it is unclear to what extent 
the two are related as material ejected in the outflow may not
participate in star formation, and in any case not before it has had time
to cool down and fall back into the galaxy.
The chemical evolution of an NGC~1569-type 
galaxy has been modelled in detail by \citet{rec06}, assuming a
variety of bursty star formation histories. Their chemo-dynamical simulations
show that elements produced in the last burst of star formation generally 
do not get mixed with the ISM in the galaxy, but instead get injected 
into the hot gas phase. None of their models produce an O abundance
as high as the one observed by us, and the model that comes closest
($12 + \log$ (O/H) $\sim 8.4$) severely underpredicts the N/O abundance
of the H{\sc ii} regions observed by \citet{ks97}. Generally, the best 
fits to the H{\sc ii} region abundances are obtained for a ``gasping'' 
star formation history,
but these models all predict a maximum $12 + \log$ (O/H) $\approx$ 8.1.
Similarly, the O abundance predicted by the models of \citet{romano06}
for NGC~1569 never exceeds $12 + \log$ (O/H) = 8.41. We speculate that
the galactic wind which is responsible for removing metals may be less
efficient in doing so near the bottom of the potential well where the
massive clusters formed.
Unfortunately, none of the models include predictions for the wide range 
of other elements we are observing here.

The Small Magellanic Cloud presents an interesting comparison case.
Early spectroscopic studies and Str{\"o}mgren photometry indicated
that the young cluster NGC~330 is about 0.5 dex more metal-poor than the 
surrounding field \citep[and references therein]{gr92}. High-dispersion 
spectroscopy by \citet{hill99} showed a smaller 
and only marginally significant difference between the field stars and 
NGC~330, although still in the sense that the cluster is more metal-poor
than the field ([Fe/H] = $-0.82\pm0.10$ vs.\
[Fe/H] = $-0.69\pm0.11$).
The [O/Fe] abundance ratios in these stars were all found to be sub-solar 
([O/Fe] $\approx$ $-$0.15 to $-$0.3 dex), as in the LMC and in young 
Galactic supergiants \citep{hbs97}.  \citet{hill99} also found the 
$\alpha$-elements abundances (Mg, Ca, Ti) relative to Fe to be around Solar 
for NGC~330.  \citet{gw99} found slightly
lower iron abundances of [Fe/H] = $-0.94\pm0.02$ for K supergiants
in NGC~330 than \citet{hill99} and an [O/Fe] ratio closer to Solar.
They derived somewhat enhanced  [Ca/Fe], [Si/Fe] and [Mg/Fe]
ratios ($+0.18$, $+0.32$ and $+0.11$), in contrast to sub-solar ratios derived 
for B stars.

While there is some evidence for a difference in metallicity between 
NGC~330 and the SMC field, the situation there seems to contrast with 
the case of NGC~1569-B where we find a \emph{higher} oxygen abundance 
of the cluster compared to the H{\sc ii} regions and an \emph{enhanced} [O/Fe] 
ratio. Our mean [$\alpha$/Fe] = $+0.31\pm0.09$ is also
higher than that derived for the stars in NGC~330. One significant difference
may be the mass of NGC~1569-B, which is probably an order of magnitude
greater than that of NGC~330 \citep{fb80}.

The spectral analysis returns a mean $T_{\rm eff} = 3800\pm200$ K and
$\log g \approx 0.0$ for the RSGs in NGC~1569-B. We have checked these 
values using resolved photometry of the RSGs in the cluster, derived
from archival HST/ACS images. About 60 RSGs are easily identifiable in
the colour-magnitude diagram, and from their
$m_{\rm F555W} - m_{\rm F814W}$ colours we derive a mean effective temperature
of $\langle T_{\rm eff}\rangle = 3850$ K and $\log g = 0.1$. An even closer 
match to the
spectroscopic $T_{\rm eff}$ is obtained if we weigh the $T_{\rm eff}$
estimates for the individual stars by their $H$-band luminosities, in
which case we get $\langle T_{\rm eff} \rangle = 3790$ K. 

We have compared the CMD with $Z=0.004$ and $Z=0.008$ isochrones from 
the Padua and Geneva groups.  Both sets of models provide the best match to 
the observed colours and magnitudes of the RSGs for $Z=0.008$ and ages of 
15--25 Myrs, but no isochrone can reproduce the observed CMD in detail. 
For these sub-solar metallicities, the 
models predict higher (by up to a few 100 K) mean effective temperatures 
$\langle T_{\rm eff} \rangle$ for the RSGs than observed.  Since the RSGs
generally become cooler with increasing metallicity, models of higher
metallicity are favoured, but models of Solar metallicity would be
required to match the observed colours. For such models, however, the
\emph{blue} supergiants become much too cool to match the observations.
The observed ratio of blue to red supergiants (BSG/RSG = $0.39\pm0.10$) is 
also significantly lower than most of the model predictions.  Although
we are detecting an impressive number of RSGs, it should be noted that
since we are only resolving stars located well outside the half-light 
radius, the actual number of RSGs in NGC~1569-B is likely to be more than 
a factor of two greater than the number we have detected.

We derive a velocity dispersion of $9.6\pm0.3$ km/s from
the integrated spectrum of NGC~1569-B, somewhat higher than the
value of 7.5 km/s of \citet{gg02}. Combining this with an estimate
of the cluster half-light radius of $0\farcs20\pm0\farcs05$ measured
on the ACS images, we obtain a dynamical mass of 
$(4.4\pm1.1)\times10^5 \left(\frac{D}{2.2 {\rm Mpc}}\right)
M_\odot$. This is in excellent agreement with the mass predicted by
simple stellar population models for a cluster of this age and
luminosity and a standard IMF.  A correction for mass segregation would
bring the two estimates into even closer agreement.

Our analysis of structural parameters reveals a decrease in core radius
with wavelength for NGC~1569-B.
This trend is consistent with the colour gradients observed 
by \citet{hunter00} and by us in the HRC data when variations in the PSF
are taken into account. A similar colour gradient was observed in
an YMC in NGC~6946 \citep{laretal01}.  The correlation between core radius 
and wavelength is also in the same sense as that found for M82-F 
by \citet{mgv05} and may be an indication that mass segregation (primordial 
or dynamical) is present in NGC~1569-B. However, this needs to be confirmed 
by a more detailed analysis, including a modelling of how mass 
segregation would translate into observables such as core- and half-light
radius.

As a final note, we find it fascinating to contemplate that only a couple of
decades ago, it was still uncertain whether NGC~1569-A and NGC~1569-B
were in fact star clusters in NGC~1569 or mere foreground stars. Today,
the Advanced Camera for Surveys on HST has made it possible to not only
settle this question definitively, but to study the individual stars
that make up these clusters.

\section*{Acknowledgements}

SSL thanks the \emph{International Space Science Institute} (ISSI) in Bern, 
Switzerland, where part of this work was done, for its hospitality. 
Mark Gieles and Evghenii Gaburov are thanked for interesting discussions
about the effect of mass segregation on integrated colours.
JPB acknowledges support from NSF grant AST-0206139.
We acknowledge the Keck Observatory and the NIRSPEC team.
The authors wish to recognise and acknowledge the very significant cultural
role and reverence that the summit of Mauna Kea has always had within
the indigenous Hawaiian community.
We are most fortunate to have the opportunity to conduct observations 
from this mountain.

\label{lastpage}

\end{document}